\documentclass[preprint2]{aastex6}
\usepackage{amsmath,natbib,apjfonts}
\pdfoutput=1
\usepackage[normalem]{ulem}

\newcommand{\ergs}{$\textmd{erg}\;\textmd{s}^{-1}$}
\newcommand{\nus}{\textit{NuSTAR}}
\newcommand{\nustar}{\hbox{\it NuSTAR}}

\newcommand{\xmm}{\hbox{\textit{XMM}-Newton}}
\newcommand{\swiftbat}{\hbox{\textit{Swift}/BAT}}
\newcommand{\swiftxrt}{\hbox{\textit{Swift}/XRT}}
\newcommand{\chandra}{\hbox{\textit{Chandra}}}
\begin{document}

\title{Hard X-ray selected AGNs in low-mass galaxies from the \textit{N\MakeLowercase{u}STAR} serendipitous survey}
\author{C.-T. J. Chen\altaffilmark{1,2}}
%\author{C.-T. J. Chen\begin{CJK*}{UTF8}{bsmi}(陳建廷)\end{CJK*}\altaffilmark{1,2}}
\author{W. N. Brandt\altaffilmark{1,2,3}}
\author{A. E. Reines\altaffilmark{4,5}}
\author{G. Lansbury\altaffilmark{6}}
\author{D. Stern\altaffilmark{7}}
\author{D. M. Alexander\altaffilmark{8}}
\author{F. Bauer\altaffilmark{9,10,11}}
\author{A. Del Moro\altaffilmark{12}}
\author{P. Gandhi\altaffilmark{13}}
\author{F. A. Harrison\altaffilmark{14}}
\author{R. C. Hickox\altaffilmark{15}}
\author{M. J. Koss\altaffilmark{16,17}}
\author{L. Lanz\altaffilmark{15}}
\author{B. Luo\altaffilmark{18}}
\author{J. R. Mullaney\altaffilmark{19}}
\author{C. Ricci\altaffilmark{9,10,20}}
\author{J. R. Trump\altaffilmark{21}}

\altaffiltext{1}{Department of Astronomy and Astrophysics, Pennsylvania State University, University Park, PA 16802, USA; ctchen@psu.edu}
\altaffiltext{2}{Institute for Gravitation and the Cosmos, Pennsylvania State University, University Park, PA 16802, USA}
\altaffiltext{3}{Department of Physics, Pennsylvania State University, University Park, PA 16802, USA}
\altaffiltext{4}{National Optical Astronomy Observatory, 950 N Cherry Ave, Tucson, AZ 85719, USA}
\altaffiltext{5}{Hubble Fellow}
\altaffiltext{6}{Institute of Astronomy, University of Cambridge, Madingley Road, Cambridge, CB3 0HA, UK}
\altaffiltext{7}{Jet Propulsion Laboratory, California Institute of Technology, Pasadena, CA 91109, USA}
\altaffiltext{8}{Centre for Extragalactic Astronomy, Department of Physics, Durham University, South Road,
  Durham, DH1 3LE, UK}
\altaffiltext{9}{Instituto de Astrof{\'{\i}}sica and Centro de Astroingenier{\'{\i}}a, Facultad de F{\'{i}}sica, Pontificia Universidad Cat{\'{o}}lica de Chile, Casilla 306, Santiago 22, Chile} 
\altaffiltext{10}{Millennium Institute of Astrophysics (MAS), Chile} 
\altaffiltext{11}{Space Science Institute, 4750 Walnut Street, Suite 205, Boulder, Colorado 80301, USA} 
\altaffiltext{12}{Max-Planck-Institut f\"{u}r Extraterrestrische Physik (MPE), Postfach 1312, D85741, Garching, Germany}
\altaffiltext{13}{Department of Physics and Astronomy, University of
Southampton, Highfield, Southampton SO17 1BJ, UK}
\altaffiltext{14}{Cahill Center for Astronomy and Astrophysics, California
Institute of Technology, Pasadena, CA 91125, USA}
\altaffiltext{15}{Department of Physics and Astronomy, Dartmouth College, 6127 Wilder Laboratory, Hanover, NH 03755, USA}
\altaffiltext{16}{Institute for Astronomy, Department of Physics, ETH
Zurich, Wolfgang-Pauli-Strasse 27, CH-8093 Zurich, Switzerland}
\altaffiltext{17}{Ambizione Fellow}
\altaffiltext{18}{School of Astronomy and Space Science, Nanjing University, Nanjing 210093, China}
\altaffiltext{19}{Department of Physics and Astronomy, The University of Sheffield, Hounsfield Road, Sheffield S3 7RH, UK}
\altaffiltext{20}{Kavli Institute for Astronomy and Astrophysics, Peking University,
Beijing 100871, China}
\altaffiltext{21}{Department of Physics, University of Connecticut, 2152 Hillside Road, Storrs, CT 06269, USA}

\shorttitle{Hard X-ray selected AGNs in low-mass galaxies}
\shortauthors{Chen et al.}
\slugcomment{ApJ accepted}

\begin{abstract}
We present a sample of 10 low-mass active galactic nuclei (AGNs) selected from the 40-month \nus\ serendipitous survey.
The sample is selected to have robust {\it NuSTAR} detections at
$3 - 24$~keV, to be at $z < 0.3$, and to have optical {\it r}-band magnitudes
at least 0.5~mag fainter than an $L_\star$ galaxy at its redshift. 
The median values of absolute magnitude, stellar mass and 2--10 X-ray luminosity of our sample are 
$\langle M_r\rangle = -20.03$, 
$\langle M_\star\rangle = 4.6\times10^{9}M_\sun$, and $\langle L_{2-10\mathrm{keV}}\rangle = 3.1\times10^{42}$ erg s$^{-1}$, respectively. 
Five objects have detectable broad H$\alpha$ emission in their optical spectra, indicating black-hole masses of $(1.1-10.4)\times 10^6 M_\sun$. 
We find that $30^{+17}_{-10}\%$ of the galaxies in our sample do not show AGN-like optical narrow emission lines, and one of the ten galaxies in our sample, J115851+4243.2, shows evidence for heavy \hbox{X-ray} absorption.
This result implies that a non-negligible fraction of low-mass galaxies might harbor accreting massive black holes that are missed by optical spectroscopic surveys and $<10$ keV \hbox{X-ray} surveys. 
The mid-IR colors of our sample also indicate these optically normal low-mass AGNs cannot be efficiently identified with typical AGN selection criteria based on {\it WISE} colors.
While the hard ($>10$ keV) \hbox{X-ray} selected low-mass AGN sample size is still limited,
our results show that sensitive {\it NuSTAR} observations are capable of probing faint hard \hbox{X-ray} emission originating
from the nuclei of low-mass galaxies out to moderate redshift ($z<0.3$),  thus providing a critical step in understanding AGN demographics in low-mass galaxies.
\end{abstract}


\section{Introduction}\label{sec:intro}
Understanding the properties of the massive black holes (mBHs) in the centers of low-mass  galaxies ($M_\star/M_\sun \la 10^{10}$) 
is an important way to discriminate observationally between different BH-seed formation scenarios \citep[e.g.,][]{volo10,gree12,rein16review}. It is also unclear whether the well-known scaling relation between the supermassive black hole (SMBH) mass and the velocity dispersion of the host-galaxy bulge extends to the low-mass regime (see \citealt[][for a review]{kh13araa}, but also see \citealt{bart05,xiao11,bald15,bald16a}). Therefore, our understanding of galaxy evolution remains incomplete without a clear picture of the mBH population in low-mass galaxies.

AGN emission powered by accretion onto mBHs in low-mass galaxies is often diluted and/or mimicked
by stellar processes in the host galaxies \citep[e.g.][]{mora02,mora14dwarf,trum15}.
Therefore, the identification of accreting mBHs in low-mass galaxies is challenging.
In practice, single-epoch spectroscopic observations at optical wavelengths have been the most-efficient method for reliably finding unobscured AGNs while also obtaining the estimated virial mass of the mBHs \citep{gree04,gh07imbh,rein13dwarf,bald15}.
However, this approach requires that AGN signatures are clearly visible (i.e., not obscured or diluted) in the optical spectrum.
Limited by the luminosity of such mBH accretion and current optical spectroscopic survey limits, 
most of the optically selected mBH candidates are unobscured AGNs at low redshifts \citep[e.g., $z\lesssim 0.35$,][]{gh07imbh,rein13dwarf}. 

On the other hand, \hbox{X-ray} stacking analyses of high-redshift galaxies in survey regions with deep \hbox{X-ray} observations have suggested that many low-mass galaxies harbor \hbox{X-ray} emitting nuclei that are heavily obscured \citep{xue12cxb,mezc15}. 
To date, there are very few known AGN residing in low-mass star-forming galaxies in the local universe (see \citealt{rein11,rein14} for individual examples). This might be due to the fact that \hbox{X-ray} observations probing energies $<10$~keV can suffer from obscuration, as well as could be affected by galaxy dilution due to the low-luminosity nature of mBH accretion \citep{rein16h210}.
Therefore, our understanding of \hbox{X-ray} selected AGN demographics in low-mass galaxies is primarily limited
to a few sources from survey regions with deep soft\footnote{Throughout the paper, we refer to X-rays win the $<10$ keV energy range as ``soft'' and $\geq 10$ keV X-rays as ``hard''.} \hbox{X-ray} observations (E-CDF-S, \citealt{schr13}; XDEEP2, \citealt{pard16}; and AMUSE, \citealt{gall08,mill15}) or archival searches \citep{lemo15}. While there are also a number of soft \hbox{X-ray} follow-up observations of mBHs selected using broad emission lines \citep[e.g.,][]{gh07xray,dong12xray,bald16b}, most of them are \hbox{X-ray} unobscured as a consequence of the optical broad-line selection. 
Furthermore, \cite{plot16} and \cite{simm16} have reported that some AGNs with broad optical emission lines in low-mass galaxies have surprisingly weak $<10$ keV X-ray emission compared to the expected values from their [\ion{O}{3}] and UV luminosities \citep{bald16b}.

An alternative approach to studying the AGN populations hosted by low-mass galaxies is to observe in hard X-rays since hard \hbox{X-ray} photons are much less susceptible to absorption by intervening material.
In addition, recent \textit{Nuclear Spectroscopic Telescope Array} \citep[\textit{NuSTAR},][]{inst_nustar}
studies of two star-forming galaxies have also revealed that the broad-band \hbox{X-ray} spectra of
galaxies powered by stellar processes are dominated by $kT\approx 0.2-1$ keV plasma emission at $E<1-3$ keV and a steep
($\Gamma\gtrsim 2.6-2.7$) power-law component at $E>5-7$ keV \citep[e.g.,][]{lehm15nustar}. 
Notably, the power-law component of some of the most luminous ULXs (ultra luminous X-ray sources) could have photon-indicies similar to those of AGNs \citep[e.g., $\Gamma \geq 1.4$,][]{walt14,mukh15}. 
Therefore, the combination of hard \hbox{X-ray} observations from \nus\ and high angular resolution data from ancillary soft \hbox{X-ray} observations can provide the means to distinguish the \hbox{X-ray} emission originating from mBHs vs. other off-nuclear stellar processes. Constructing a {\it NuSTAR}-selected AGN sample hosted by low-mass galaxies could provide a critical step to understanding the low-mass AGN population. \par

Low-mass AGNs detected by previous hard \hbox{X-ray} observatories are scarce due to the limited sensitivity and angular resolution of the previous generation of instruments. 
Even with the all-sky coverage of the \swiftbat\ survey \citep[e.g.,][]{koss11bathost}, 
the number of low-mass AGNs is small, and the low-mass AGNs detected by \swiftbat\ primarily are comprised of nearby ($z<0.005$) or luminous ($L_{14-195 \mathrm{keV}}>10^{43}$ erg s$^{-1}$) sources that might not be representative of low-mass AGNs generally. The recently launched \nus\ observatory provides a $> 100$ times improvement in hard \hbox{X-ray} sensitivity over previous observations at $\gtrsim 10$ keV. One of the first ten \nustar\  detections in the \nustar\
serendipitous survey \citep{alex13nustar} has already been identified as an AGN hosted by a dwarf galaxy ($M_\star\approx 1.5\times10^9M_\sun$). In this work, we report the properties of low-mass galaxies detected in the \nustar\ serendipitous-survey catalog \citep[][L17 hereafter]{lans17ser}, aiming to improve our understanding of the hard \hbox{X-ray} emitting AGN population hosted by low-mass galaxies.

This paper is organized as follows: in \S2, we describe the selection of low-mass galaxies from the \nus\ serendipitous survey. The data analysis is presented in \S3.
In \S4, we study the AGN properties of our sample.
In \S5, we compare the multiwavelength properties of our \nus\ sample with those of previous AGN samples hosted by low-mass galaxies. A discussion and summary are provided in \S6. Detailed data analysis for individual objects is presented in Appendix A. Throughout the paper, we assume $H_0=70$ km s$^{-1}$Mpc$^{-1}$ and a $\Lambda$CDM cosmology with $\Omega_m=0.3$ and $\Omega_\Lambda=0.7$. The uncertainties reported in this work are $1\sigma$ unless stated otherwise.

\section{Sample selection and stellar-mass estimation}
\begin{figure}
\epsscale{1.15}
\plotone{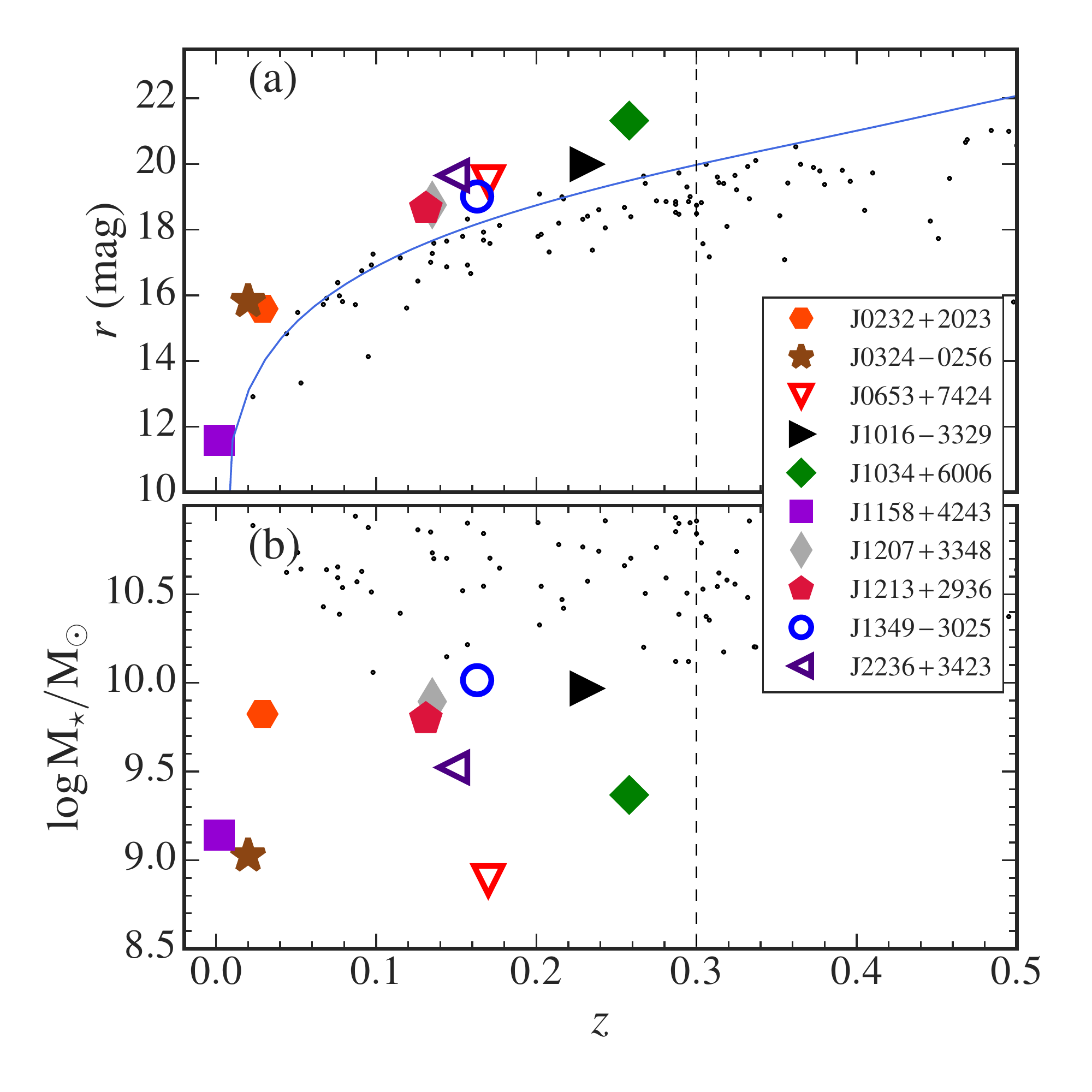}
\vspace{-0.5cm}
\caption{Redshift as a function of (a) {\it r}-band AB magnitudes  and (b) $M_\star$ for the $z<0.3$ \nus\ low-mass AGN sample. The dashed line marks the $z=0.3$ redshift cut.
Our sample is selected to be at least 0.5 mag fainter than an $L_\star$ galaxy at the corresponding redshift. The observed-frame {\it r}-band luminosity of an $L_\star$ galaxy derived from the \cite{kelv14} luminosity function is plotted as the solid line in (a). The full spectroscopic sample at $z<0.5$ from the \nus\ serendipitous survey is shown as gray dots.}
\label{fig:rmag_mass}
\end{figure}

\subsection{The \textit{NuSTAR} serendipitous survey}\label{subsec:nus_ser}
To date, the \nus\ serendipitous survey (L17) has covered an area of $\approx$13 deg$^{2}$ and detected $497$ individual sources. The serendipitous survey is the largest-area component of the \hbox{{\it NuSTAR}} extragalactic survey program. This program searches for serendipitous detections in most of the \nus\ pointings.
In short, the serendipitous catalog includes soft band ({\it NuSTAR}-SB, 3--8 keV) and hard band ({\it NuSTAR}-HB, 8--24 keV) photometry for sources detected in the full band ({\it NuSTAR}-FB, 3--24 keV). Sources were extracted using the source-detection software package SExtractor \citep{soft_sextract} on the false-positive probability ($P_{\rm False}$ hereafter) map generated from the mosaics of the two \nus\ focal-plane modules (FPMA and FPMB).
A more detailed discussion of source detection and the $P_{\rm False}$ map generation can be found in \cite{mull15nustar} and L17. \par

In this work, we focus on the 248 objects that are spectroscopically confirmed to be extragalactic.
Their redshifts were primarily obtained from the dedicated spectroscopic
follow-up observations as part of the \nus\ serendipitous survey (see L17 for details), and also from publicly available spectroscopic surveys. We refer readers to L17 for details and a complete discussion of \nus\ source-extraction methods and the \nus\ serendipitous catalog.

\subsection{Multiwavelength counterpart identification}\label{subsec:phot}
Source matching and counterpart identification for \nus\ are challenging due to the relatively large PSF ($\approx 18\arcsec$ FWHM) and astrometric uncertainty \citep[$\approx 6.6\arcsec$, see][]{civa15nustar}. For faint sources, the positional uncertainty of \nus\ can be as high as $\approx 22\arcsec$ (see \S3.1 of L17).
We first match each \nus\ source position to the closest {\it Chandra}, {\it XMM}-Newton,
or \swiftxrt\ source when available, as the positional accuracy of these lower-energy \hbox{X-ray} observatories is better. 
We then match the soft X-ray positions to the Sloan Digital Sky Survey \citep[SDSS,][]{sdss_main} or the SuperCOSMOS scans of photographic Schmidt plates \citep{cat_supercosmos} with a $5\arcsec$ search radius.
We note that 15 of the 248 sources do not have ancillary soft \hbox{X-ray} observations. 
For these objects, we use a searching radius of $22\arcsec$ around the \nus\ positions to search for their optical counterparts. Based on visual inspection of the optical images, we reject seven of the 15 objects with multiple optical and infrared counterparts within the search radius to avoid potentially spurious matching results. Each of the other eight sources is matched to the only one optical counterpart within the $22\arcsec$ search radius. We note that none of the 15 objects is included in our final sample of low-mass AGNs.

Of the 248 extragalactic \nus\ serendipitous sources, all have {\it r}-band photometry.
113 of the 248 objects are covered in the imaging footprint of SDSS,
and all of these objects have optical photometry in the {\it u, g, r, i, z} bands.
To measure accurately the host-galaxy color and luminosity,
we adopt the extinction-corrected model magnitudes that are scaled
to the {\it i}-band c-model magnitude\footnote{\url{http://www.sdss.org/dr12/spectro/galaxy_portsmouth/}} from SDSS DR12 \citep{sdss_dr12}.

For the 135 objects outside of the SDSS imaging footprint, the optical photometry was obtained from SuperCOSMOS, which is considered to have 0.3 mag photometric uncertainties. \par

Near-IR constraints for our sample come from the 
{\it J, H}, and {\it Ks} band photometric catalog of the Two Micron All Sky Survey \citep[2MASS,][]{cat_2mass}. We also obtain mid-IR photometry from the ALLWISE catalog, which is an all-sky catalog covering the $3.4, 4.6, 12$, and $22\micron$ bands ({\it W1, W2, W3}, and {\it W4} hereafter) observed with the {\it Wide Field Infrared Survey Explorer} \citep[\textit{WISE},][]{cat_wise}.
We make use of the UV {\it GALEX} \citep{inst_galex} photometry from the {\it GALEX} Release 6/7 \citep{cat_galex67}.
We correct for the Galactic extinction in the {\it GALEX} near- and far-UV bands using the {\it E(B-V)} values from the {\it GALEX} catalog and
the $R_V$ values from \cite{wyde07galex}. For the optical, near-IR, and mid-IR bands,
the Galactic extinction values were obtained using the IRSA Galactic Reddening and Extinction Calculator using the \cite{schl98ext} extinction map.
Counterparts in the UV, near-IR, and mid-IR bands were obtained by searching around the optical positions within a $5\arcsec$ radius.
We note that three of the low-redshift serendipitous objects appear to be extended in the near-IR and mid-IR images (J023229+2023.7, J032459--0256.2, and J115851+4243.2, see Figure set~\ref{fig:figset}) and the default profile-fit photometry provided by 2MASS and ALLWISE might not be optimal.
To obtain accurate photometry for these three extended sources, we adopt the $20$ mag/deg$^2$ isophotal fiducial elliptical magnitudes in the 2MASS {\it J, H}, and {\it Ks} bands and the {\it WISE} magnitudes measured via elliptical aperture photometry. 
All 248 objects have a signal-to-noise ratio (SNR) $>5$ in the {\it WISE W1} and {\it W2} bands.
With the more limited sensitivity of 2MASS and {\it GALEX}, only 80 objects are detected in the {\it J, H}, and {\it Ks} bands,
and only 114 objects are detected by {\it GALEX}.
\par

\begin{deluxetable}{c|cllLl}
\tabletypesize{\scriptsize}
\tablecolumns{6}
\tablewidth{0pt}
\tablecaption{\nus\ Low-Mass Galaxies}\label{tab:sample}
\tablehead{
\colhead{ID} &
\colhead{Source Name} &
\colhead{Target Field} &
\colhead{RA (J2000)} &
\colhead{DEC (J2000)} &
\colhead{z}}
\colnumbers
\decimals
\startdata
1 & J023229+2023.7 & 1ES 0229+200 & 38.120106 &  20.396729 & 0.029\\
2 & J032459--0256.2 & NGC 1320 & 51.249614 &  -2.9364875 & 0.020\\
3 & J065318+7424.8 & Mrk 6 & 103.331370 & 74.418452 & 0.170\\
4 & J101609--3329.6 & IC2560 & 154.003625 & -33.493796 & 0.231\\    
5 & J103410+6006.7 & Mrk 34$^a$ & 158.541966 & 60.112053 & 0.258\\
6 & J115851+4243.2$^b$ & IC 751 & 179.713759  & 42.721351 & 0.002\\
7 & J120711+3348.5 & B2 1204+34  & 181.797028 &  33.807852 & 0.135\\
8 & J121358+2936.1 & WAS 49b & 183.494820 &  29.602344 &  0.131\\
9 & J134934--3025.5 & IC 4329A & 207.39206 &  -30.427494 & 0.163\\
10 & J223654+3423.5 & SN 2014C & 339.226251 &  34.391265 & 0.148\\
\enddata
\tablecomments{Column 1 : ID number (for referencing the online figure set~\ref{fig:figset}). 
Column 2 : Source name used in the \nus\ serendipitous catalog (L17). 
Column 3 : The science target field associated with the serendipitous object.
Columns 4--5 : \nus\ FPMA+FPMB RA and DEC (J2000) of the serendipitous source. 
Column 6 : Redshift.\\
$^a$ : Also known as SDSS J1034+6001 in \cite{gand14}. \\
$^b$ : Also known as IC 750.}
\end{deluxetable}
\vspace{-1.2cm}

\subsection{The NuSTAR low-mass galaxy sample}
To construct a sample of low-mass galaxies with robust detections in hard X-rays,
we focus on a low-redshift ($z<0.3$) subsample of the \nus\ serendipitous-survey sources covered by both FPMA and FPMB and having $\mid b \mid>10\degr$. The redshift cut is motivated by the sensitivity limit of \textit{NuSTAR}.
Beyond $z\approx 0.3$, even for an mBH with $M_\bullet=10^{7}M_\sun$ radiating at the Eddington limit ($L_{10-40\mathrm{keV}} \approx 6\times10^{43}$ erg s$^{-1}$), the expected count rates in the {\it NuSTAR}-HB ($8-24$ keV) would still be too low for \nus\ to provide detections in the typical exposure times of the \nus\ serendipitous survey (the serendipitous survey has a median exposure time of 28 ks, see L17).

We next search for candidate low-mass galaxies by comparing the extinction-corrected {\it r}-band magnitude (observed-frame AB magnitudes) with that of an $L_\star$ galaxy (i.e., a galaxy with a luminosity equal to the value of the ``knee'' of the luminosity function) calculated using the \cite{kelv14} luminosity function for elliptical galaxies. There are a total of 10 sources at $z<0.3$ with an $r$-band magnitude that is at least 0.5 mag fainter than $r_\star$ ($r$-band magnitude of an $L_\star$ galaxy) at the corresponding redshift. \par

L17 report a spurious matching rate of $\approx 7\%$ between the \nus\ and soft X-ray positions. To ensure that the \nus\ positions of the low-mass galaxy candidates are matched to the correct counterpart, we visually inspected the soft X-ray, optical, near-IR, and mid-IR images for each of the 10 objects.
We find that 9 of the 10 sources have only one soft \hbox{X-ray} counterpart within a $22\arcsec$ radius. 
The other object, J115851+4243.2, has three soft X-ray point sources within a $22\arcsec$ radius, 
but only the source at the galactic center has a \hbox{3--8 keV} flux comparable to the \nus\ \hbox{3--8 keV} flux. We adopt the central source as the {\it NuSTAR} counterpart. Therefore, we consider the matching between the \nus\ positions and the soft X-ray positions to be accurate. For each source, the soft \hbox{X-ray} position has only one optical counterpart within a $5\arcsec$ search radius.

As a result, we have a final sample of 10 low-mass galaxy candidates with \nus\ detections.
This sample-selection approach is chosen because the stellar-mass measurements are often sensitive to the choice of initial mass function and the uncertainty of stellar population synthesis models used to estimate $M_\star$. 
For AGN host galaxies, the stellar-mass estimation is further complicated by the AGN emission. We note that, by happenstance, our empirical selection criterion of {\it r}$< r_\star-0.5$ recovers all of the $M_\star<10^{10}M_\sun$ galaxies detected in the \nus\ serendipitous catalog at $z<0.3$ (see \S\ref{subsec:sedfitting} for the details of stellar-mass estimation for the \nus\ serendipitous catalog). 
We show the redshift versus {\it r}-band magnitude distribution for our low-mass galaxy sample in Figure~\ref{fig:rmag_mass}a.

All 10 galaxies in our sample satisfy the $\log P_{FALSE} < -6$ significance criterion in the $3-24$ keV band (L17).
In {\it NuSTAR}-HB, only two objects in our sample pass the same false-detection probability criterion.
Since the objects in our sample were carefully matched to their soft \hbox{X-ray} to mid-IR counterparts,
it is less likely for the \nus\ detections at these positions to be caused by random Poisson noise.
Thus, we consider the eight objects with $>2\sigma$ net counts in the $8-24$ keV band to be reliably detected in {\it NuSTAR}-HB. For the other two objects with {\it NuSTAR}-HB net counts less than $2\sigma$, we consider these objects as non-detected. We adopt the gross source counts (background plus source counts) uncertainty estimated using the \cite{gehr86} method as the HB net counts upper limits for these objects without HB detections. 
The \nus\ low-mass AGNs are presented in Table~1. Their \nus\ photometric properties are listed in Table~\ref{tab:nusprop}, and their multiwavelength properties are listed in Table~\ref{tab:mwprop}. 
Since the \nus\ positional uncertainty is $\sim 22\arcsec$, we also list the angular offset between the \nus\ positions and the optical positions in Table~\ref{tab:mwprop}.

\newpage
\floattable
\begin{deluxetable*}{lcccccccccccc}
\tabletypesize{\footnotesize}
\rotate
\tablewidth{0pt}
\tablecaption{\nus\ Source Properties\label{tab:nusprop}}
\tablehead{
\colhead{Source Name} &
\colhead{OBSID} &
\colhead{$\log P_\mathrm{False}$} &
\colhead{Exp} &
\colhead{Flux} &
\colhead{Flux} &
\colhead{Flux} &
\colhead{Net source counts} &
\colhead{Net source counts} &
\colhead{Background counts} &
\colhead{Band ratio} &
\colhead{$\log L_{10-40\mathrm{keV}}$} & 
\colhead{$\log L_{10-40\mathrm{keV}}^{\rm Model}$} \\
\colhead{} & \colhead{} & \colhead{} & \colhead{(ks)} & \colhead{(3--8 keV)} & \colhead{(8--24 keV)} &
\colhead{(3--24 keV)} & \colhead{(3--8 keV)} & \colhead{(8--24 keV)} & \colhead{(3--24 keV)}& \colhead{} & \colhead{(erg s$^{-1}$)}
& \colhead{(erg s$^{-1}$)}}
\colnumbers
\startdata
\vspace{5pt}
J023229+2023.7 & 60002047 &-83.1 & 37.5 & $40.33\pm3.08$ & $71.62\pm7.29$ & $117.08\pm7.1$ & $275\pm21$ & $193\pm20$ & 279 & $0.68^{+0.07}_{-0.07}$ & 42.3 & 42.6 (42.6)\\
\vspace{5pt}
J032459--0256.2 & 60061036 & -49.8 & 12.0  & $36.07\pm5.08$ & $49.78\pm10.55$ & $94.6\pm10.81$ & $78\pm11$ & $43\pm9$ & 42 & $0.54^{+0.1}_{-0.11}$ &41.8 & 41.7 (41.7)\\
\vspace{5pt}
J065318+7424.8 & 60102044 & -9.9 & 71.4  & $2.84\pm1.03$ & $6.14\pm2.46$ & $8.59\pm2.23$ & $34\pm12$ & $30\pm12$ & 187 &$0.86^{+0.28}_{-0.33}$ & 42.8 & 43.1 (43.1)\\
\vspace{5pt}
J101609--3329.6 & 50001039 & -15.0$^a$ & 43.8  & $10.91\pm2.37$ & $12.07\pm5.4$ & $27.0\pm5.46$ & $76\pm16$ & $31\pm14$ & 146 & $0.35^{+0.08}_{-0.11}$ & 43.4 & 43.7 (43.7)\\
\vspace{5pt}
J103410+6006.7 & 60001134 & -9.1 & 25.0  & $4.67\pm1.94$ & $12.81\pm4.85$ & $16.26\pm4.38$ & $20\pm8$ & $23\pm9$ & 70 &$1.15^{+0.35}_{-0.46}$ &43.4 & 42.9 (42.9) \\
\vspace{5pt}
J115851+4243.2 & 60001148 & -8.3 & 39.0  & $5.34\pm1.56$ & $7.88\pm3.99$ & $14.63\pm3.71$ & $38\pm11$ & $22\pm11$ & 145 &$0.45^{+0.14}_{-0.21}$ &39.0 & 38.7 (38.8)\\
\vspace{5pt}
J120711+3348.5 & 60061356 & -6.1 & 14.8  & $10.07\pm3.63$ & $<12.29^b$ & $25.53\pm7.72$ & $18\pm7$ & $<9^b$ & 23 & $0.44^{+0.16}_{-0.23}$ & $<43.1^b$ & \nodata \\
\vspace{5pt}
J121358+2936.1 & 60061335 &-14.9 & 16.1  & $10.46\pm2.91$ & $16.74\pm6.71$ & $29.24\pm6.48$ & $31\pm9$ & $20\pm10$ & 51 & $0.61^{+0.18}_{-0.21}$ & 43.0 & 42.2 (42.2)\\
\vspace{5pt}
J134934--3025.5 & 60001045 & -9.8 & 125.4  & $2.85\pm0.86$ & $6.91\pm2.11$ & $9.57\pm2.01$ & $65\pm19$ & $62\pm19$ & 534 & $0.99^{+0.26}_{-0.35}$ & 42.7 & \nodata \\
\vspace{5pt}
J223654+3423.5 & 80001085 & -2.7$^a$ & 79.7  & $2.47\pm1.3$ & $<7.57^b$ & $6.16\pm2.89$ & $27\pm14$ & $<35^b$ & 189 & $1.3^{+0.42}_{-0.53}$ & $<42.8^b$ & \nodata \\
\enddata
\tablecomments{Column 1 : Source name. \\
Column 2 : \nus\ observation ID. \\
Column 3 : {\tt SExtractor} false-detection probability for the $3-24$ keV band, see \S\ref{subsec:nus_ser}.\\
Column 4 : Nominal exposure time of the primary science target.\\
Columns 5--7 : Aperture-corrected fluxes in the 3--8 keV (SB), 8--24 keV (HB), and 3--24 keV (FB) energy bands in units of $10^{-14}$ erg s$^{-1}$ cm$^{-2}$. The fluxes were calculated assuming a power-law with photon index $\Gamma=1.8$, see L17 for details.\\
Columns 8--9 : Background-subtracted photon counts in the 3--8 keV and 8--24 keV energy bands.\\
Column 10 : Background photon counts in the 3--24 keV band.\\
Column 11 :  HB source counts/SB source counts, calculated using Bayesian Estimation of Hardness Ratios method \citep[BEHR,][]{soft_behr}.\\
Column 12 : Rest-frame 10--40 keV luminosity calculated based on the 8--24 keV fluxes and a $\Gamma=1.8$ power-law spectrum (not corrected for absorption).\\
Column 13 : Rest-frame 10--40 keV luminosity calculated based on the best-fit models discussed in \S\ref{subsec:softxray}. Absorption-corrected luminosities are given in parentheses. For targets with quality \nus\ spectra, the luminosities are derived from the \nus\ component of the joint-fit. See Appendix A and Table 4 for details on the model and data used for each source.
\\
$^a$ : Sources are from the ``secondary catalog'' of the \nus\ serendipitous survey. The secondary catalog is constructed using a different source-detection method than the primary catalog. For these objects, the probability reported here is the binomial no-source probability; see L17 for details. \\
$^b$ : $1\sigma$ upper limits.}
\end{deluxetable*}

\newpage
\floattable
\begin{deluxetable*}{llllllllllhl}
\tabletypesize{\scriptsize}
\rotate
\tablecolumns{12}
\tablewidth{0pt}
\tablecaption{\textit{NuSTAR} Source Photometry \label{tab:mwprop}}
\tablehead{\colhead{Name$^a$} & \colhead{J023229+2023.7} & \colhead{J032459--0256.2} & \colhead{J065318+7424.8} & \colhead{J101609--3329.6} & \colhead{J103410+6006.7} & \colhead{J115851+4243.2} &
\colhead{J120711+3348.5} & \colhead{J121358+2936.1} & \colhead{J134934--3025.5} &  & \colhead{J223654+3423.5}}
\startdata     
RA  (J2000)$^b$     & 38.120017 & 51.249724 & 103.337760 & 154.00501 & 158.543011 & 179.71751 & 181.796853 & 183.497104 & 207.388010 & 331.641886 & 339.224396 \\
DEC (J2000)$^b$     & 20.397181 & -2.936738 & 74.417266 & -33.492958  & 60.112193  & 42.72247  & 33.811587 & 29.604993 & -30.426759 & 10.194054 & 34.392549 \\
Optical Offiset$^c$ & 1.6 & 1.0  & 7.5 & 5.1 & 1.9 &  10.7 &  13.4 &  11.7 &  12.8 & \nodata & 7.3 \\
{\it GALEX} Far-UV  & \nodata & $19.06\pm0.12$ & \nodata\ & \nodata & $22.90\pm0.13$  & $17.85\pm0.07$ & $20.75\pm0.21$ & $20.95\pm0.10$ & $23.28\pm0.32$  & \nodata & \nodata \\
{\it GALEX} Near-UV & \nodata & $18.60\pm0.06$ & \nodata\ & \nodata & \nodata         & \nodata        & $20.74\pm0.13$ & $20.54\pm0.05$ & $22.50\pm0.12$  & \nodata & $23.08\pm0.26$ \\
{\it u}$^d$ & $17.84\pm0.03$ & \nodata & \nodata & \nodata & $23.49\pm0.69$ & $14.64\pm0.01$ & $20.08\pm0.04$ & $19.7\pm0.04$ & \nodata & $24.39\pm0.97$ & $21.47\pm0.37$ \\
{\it g/B}$^d$ & $16.38\pm0.01$ & $16.59$ & $21.57$ & \nodata & $22.27\pm0.1$ & $12.57\pm0.01$ & $19.44\pm0.01$ & $19.16\pm0.01$ & $20.3$ & $22.79\pm0.2$ & $20.54\pm0.04$ \\
{\it r/R}$^d$ & $15.59\pm0.01$ & $15.95$ & $19.77$ & $20.2$ & $21.32\pm0.08$ & $11.58\pm0.01$ & $18.75\pm0.01$ & $18.65\pm0.01$ & $19.13$ & $22.81\pm0.31$ & $19.65\pm0.03$ \\
{\it i/I}$^d$ & $15.21\pm0.01$ & $15.41$ & \nodata & $19.02$ & $20.96\pm0.08$ & $11.06\pm0.01$ & $18.32\pm0.01$ & $18.18\pm0.01$ & $18.44$ & $22.14\pm0.25$ & $19.19\pm0.03$ \\
{\it z}$^d$ & $14.93\pm0.01$ & \nodata & \nodata & \nodata & $20.94\pm0.34$ & $10.62\pm0.01$ & $18.08\pm0.02$ & $18.11\pm0.04$ & \nodata & $21.97\pm0.7$ & $18.59\pm0.08$ \\
{\it J}$^d$ & $13.89\pm0.05$ & $14.23\pm0.06$ & \nodata & \nodata & \nodata & $9.28\pm0.01$ & $16.85\pm0.16$ & $17.02\pm0.22$ & \nodata & \nodata & \nodata \\
{\it H}$^d$ & $13.18\pm0.06$ & $13.63\pm0.08$ & \nodata & \nodata & \nodata & $8.5\pm0.01$ & $15.87\pm0.16$ & $15.91\pm0.16$ & \nodata & \nodata & \nodata \\
{\it Ks}$^d$ & $12.58\pm0.06$ & $13.3\pm0.1$ & \nodata & \nodata & \nodata & $8.15\pm0.01$ & $15.63\pm0.22$ & $15.42\pm0.19$ & \nodata & \nodata & \nodata \\
{\it WISE W1}$^d$ & $11.67\pm0.02$ & $13.2\pm0.02$ & $15.76\pm0.05$ & $15.55\pm0.04$ & $17.08\pm0.11$ & $7.88\pm0.0$ & $14.67\pm0.03$ & $14.93\pm0.03$ & $14.88\pm0.03$ & $14.9\pm0.03$ & $15.47\pm0.04$ \\
{\it WISE W2}$^d$ & $10.8\pm0.02$ & $13.04\pm0.03$ & $14.33\pm0.04$ & $14.8\pm0.06$ & $16.75\pm0.26$ & $7.66\pm0.0$ & $13.82\pm0.04$ & $14.03\pm0.04$ & $14.3\pm0.05$ & $13.56\pm0.03$ & $14.81\pm0.06$ \\
{\it WISE W3}$^d$ & $7.74\pm0.02$ & $9.86\pm0.05$ & $11.25\pm0.16$ & $12.11\pm0.28$ & $12.8$ & $3.94\pm0.01$ & $11.83\pm0.27$ & $10.79\pm0.1$ & $11.08\pm0.1$ & $10.48\pm0.08$ & $11.34\pm0.17$ \\
{\it WISE W4}$^d$ & $5.46\pm0.04$ & $7.7\pm0.16$ & $8.23\pm0.24$ & $8.9\pm0.38$ & $8.75$ & $1.83\pm0.01$ & $8.76\pm0.4$ & $8.72\pm0.49$ & $9.11\pm0.46$ & $7.89\pm0.19$ & $8.66\pm0.39$ \\
$M_r$$^e$ & $-$19.95 & $-$18.89 & $-$20.12 & $-$20.34 & $-$19.28 & $-$18.11 & $-$20.29 & $-$20.32 & $-$20.48 & $-$19.61 & $-$19.61 \\
Telescope$^f$ & Keck        & Keck        & Keck          & NTT        & Keck       & SDSS & SDSS & SDSS & NTT        & Keck & Keck \\
Camera$^f$    & LRIS        & LRIS        & LRIS          & EFOSC2     & DEIMOS      & \nodata   & \nodata   & \nodata   & EFOSC2     & LRIS & LRIS \\
UT date$^f$   & 2013 Oct 03 & 2013 Nov 09 & 2015 Dec 05   & 2015 Mar 14& 2013 Dec 11& 2003 Apr 25   & 2004 Apr 16   & 2004 Dec 13   & 2015 Mar 14& 2013 Oct 04 & 2016 Aug 06 \\
%Redshift$^g$  & 0.029       & 0.020 & 0.170 & 0.231 & 0.258 & 0.002 & 0.135 & 0.131 & 0.163 & 0.291 & 0.148 \\
Type$^g$      & BLAGN   & BLAGN & Galaxy & NLAGN & BLAGN & NLAGN & BLAGN & BLAGN & Galaxy & BLAGN & Galaxy \\
$\log L_{6\micron}$(AGN)$^h$ & 42.99 & 41.2 & 43.31 & 43.37 & 42.83 & 41.99 & 43.19 & 42.96 & 43.18 & 44.13 & 42.96 \\
$M_\star$$^i$ & 9.8 & 9.0 & 8.9 & 10.0 & 9.4 & 9.1 & 9.9 & 9.8 & 10.0 & 8.6 & 9.5 \\
$M_\bullet$$^j$               & 7.02 & 6.06 &  \nodata & \nodata & 7.5 &  \nodata & 6.51 & 6.82 & \nodata &  \nodata & \nodata \\
$\lambda_{\rm Eddington}$$^k$ & 6.2  & 5.3  & \nodata  & \nodata & 6.5 & \nodata  & 24.5 & 10.6 & \nodata &  \nodata & \nodata \\
\enddata
\tablecomments{
$^a$ : Source name. \\
$^b$ : Optical counterpart source position. \\
$^c$ : Offset between {\it NuSTAR} and optical source positions in arcsec.\\
$^d$ : Source photometry in its original form, e.g. AB asinh mag for SDSS and Vega for SuperCOSMOS, 2MASS and {\it WISE} bands. See \S\ref{subsec:phot} for details.\\
$^e$ : Absolute {\it r}-band magnitude.\\
$^f$ : Observational details of the optical spectroscopy.\\
%$^g$ : Optical spectroscopic redshift.\\
$^g$ : Optical spectroscopic classification. See \S\ref{subsec:optspec} for details of the classification. \\
$^h$ : Mid-IR luminosity of the AGN component from the best-fitting SED in logarithmic erg s$^{-1}$.\\
$^i$ : Stellar mass in logarithmic $M_\sun$. The typical uncertainty in $M_\star$ is 0.3 dex (see \S\ref{subsec:sedfitting})\\
$^j$, $^k$ : Black hole mass in logarithmic $M_\sun$ and Eddington ratio in percentage for broad-line AGNs. See \S\ref{subsec:mbh} for details.}
\end{deluxetable*}

\newpage
\floattable
\begin{deluxetable*}{clcccccCcccccc}%14 columns
\tabletypesize{\scriptsize}
\rotate
\tablewidth{0pt}
\tablecaption{Additional X-ray observations and X-ray spectral properties\label{tab:sxprop}}
\tablehead{
\colhead{Source Name} & %1
\colhead{Observatory} &  %2
\colhead{ObsID} & %3
\colhead{RA} & %4
\colhead{DEC} & %5
\colhead{Offset} &  %6
\colhead{Exposure time} & %7
\colhead{Flux} & %8
\colhead{Energy range} &  %9
\colhead{$\Gamma$} &  %10
\colhead{$\log N_{\rm H}^{\rm Galactic}$} & %11
\colhead{$\log N_{\rm H}$} & %12
\colhead{{\it Goodness}} & %13
\colhead{$\log L_{2-10\mathrm{keV}}$} \\ %14
\colhead{} & \colhead{(Soft X-ray)} & \colhead{} & \colhead{(J2000)} & \colhead{(J2000)} & \colhead{(\arcsec)} & \colhead{(ks)} &
\colhead{(3--8 keV)} & \colhead{(keV)} & \colhead{} & \colhead{(cm$^{-2}$)} & \colhead{(cm$^{-2}$)} & \colhead{(\%)} & \colhead{(erg s$^{-1}$)}
}
\colnumbers
\startdata
J023229+2023.7 & {\it NuSTAR}$^a$ & \nodata & \nodata & \nodata & \nodata & 37.5   & 70.47^{+9.70}_{-9.52} & 3--20 & 1.8 & 20.38 & $23.11^{+0.21}_{-0.28}$ & 32.00 & 42.25 (42.56)\\
J032459--0256.2 & \swiftxrt\ + \nus\   & 0405240201 &51.24934  & -02.93677 & $1.38$  & 16.4   & 36.5^{+6.1}_{-5.2}      & 0.5--32  & $1.94^{+0.34}_{-0.15}$ & 20.73 & $< 21.53$ & 21.40   & 41.52 (41.53) \\
J065318+7424.8 & \chandra\ + \nus\    & 10324      & 103.33762 & 74.417292 & $0.16$ & 74.1  &  2.6^{+0.2}_{-0.2} & 0.5--32 & $1.88^{+0.29}_{-0.27}$ & 20.64 & $21.32^{+0.27}_{-0.25}$ & 16.24 & 42.52 (42.54)\\
J101609--3329.6 & \xmm\ + \nus\      & 0203890101 &154.00490 & -33.49289 & $0.24$ & 71.5    & 6.7^{+0.7}_{-0.7} & 2--24 & $1.80^{+0.42}_{-0.41}$ & 20.98 & $22.79^{+0.17}_{-0.25}$ & 20.40 & 43.13 (43.27) \\
J103410+6006.7 & \xmm\         & 0306050701 &158.54270 & 60.11219  & $0.53$ & 8.8   & 3.91^{+1.0}_{-0.7}    & 0.5--12 & $1.67^{+0.18}_{-0.17}$ & 20.20 & $<20.95$  & 17.40 & 43.06 (43.06)\\
J115851+4243.2 & {\it Chandra} & 17006 & 179.71799 & 42.722333  & $1.2$ &29.6     & 1.5^{+0.5}_{-0.3} & 1.0--8 & $1.8$ & 20.17 & $23.06^{+0.35}_{-2.5}$ & 0.10	 & 38.43 (38.73) \\
J120711+3348.5 & \swiftxrt\  & 37315      & 181.79622        & 33.812419 & $3.6$ & 11.1             & 6.3\pm1.8   & 0.2--10 & 1.8 & 20.07 & $<20.07$ $^b$ & \nodata & 42.65 (42.65$^b$)\\
J121358+2936.1 & \chandra\     & 14042      & 183.497154       & 29.60475        & $0.7$ & 5.0      & 4.43^{+1.5}_{-1.4}  & 0.5--8 & $2.56^{+2.12}_{-1.25}$ & 20.34 & $21.72^{+0.4}_{-1.7}$ & 0.01 & 42.56 (42.60)\\
J134934--3025.5 & \xmm\         & 0101040401 &207.38800 & -30.42731 & $2.0$  & 9.2   & 2.3\pm0.5 & 0.2--12 & 1.8 & 20.93 & $22.0^{+0.5}_{-0.7}$ $^b$ & \nodata &    42.46 (42.50$^b$)\\
J223654+3423.5 & \chandra\  & 17570      & 339.22414      & 34.392484 & $0.7$  & 9.9    & 0.4\pm0.17  & 0.5--7 & 1.8 & 21.05 & $22.0^{+0.2}_{-0.6}$ $^b$ & \nodata  & 41.59 (41.63$^b$)\\
\cutinhead{Additional Soft X-ray Observations$^c$}
J023229+2023.7a & \xmm\         & 0604210201 &38.11999  & 20.39719  & $0.07$   & 8.7   & 43.4      & 0.2--12 & 1.8 & 20.38 & \nodata & \nodata & \nodata \\%1 = 1ES0229p200
J023229+2023.7b & \xmm\         & 0604210301 &38.11995  & 20.39723  & $0.25$   & 10.0   & 27.2      & 0.2--12 & 1.8 & 20.38 & \nodata & \nodata & \nodata \\
J032459--0256.2a & \xmm\         & 0405240201 &51.24934  & -02.93677 & $0.48$  & 11.8   & 10.2      & 0.2--12 & 1.8  & 20.73 & \nodata & \nodata & \nodata\\
J065318+7424.8a & \xmm\         & 0061540101 & 103.33733 & 74.41740  & $0.64$ & 7.1   &  5.1  & 0.2--12 & 1.8 & 20.64 & \nodata & \nodata & \nodata\\
J065318+7424.8b & \xmm\         & 0144230101 & 103.33738 & 74.41739  & $0.58$ & 29.2  & 2.0   & 0.2--12 & 1.8 & 20.64 & \nodata & \nodata & \nodata \\
J101609--3329.6a & \chandra\        & 0203890101 &154.00490 & -33.49289 & $0.25$ & 7.2    & 7.3 & 0.5--8 & 1.8 & 20.98 & \nodata & \nodata & \nodata\\
J115851+4243.2a & \xmm\         & 0744040301     & \nodata  & \nodata   & $3.84$ & 18.0  & 2.68 & 0.2--12 & 1.8 & 20.17 & \nodata & \nodata & \nodata \\
J223654+3423.5a & \chandra\ & 17569 & 339.224 & 34.393341 & $3.05$ & 9.9  & 0.4 & 0.5--8 & 1.8 & 21.05 & \nodata & \nodata & \nodata \\
J223654+3423.5b & \chandra\ & 17571 & 339.22437 & 34.393573 & $3.69$ & 9.9  & 0.2 & 0.5--8 & 1.8 & 21.05 & \nodata & \nodata & \nodata \\
\enddata
\tablecomments{Column 1 : Source name.
Additional soft \hbox{X-ray} observations for objects with data from more than one ancillary observatories are presented in the second part of this table.\\
Column 2 : Source of ancillary soft-X-ray observations.\\
Column 3 : Observation ID, ObsID, Target ID, Event file ID for {\it NuSTAR}, {\it Chandra}, \textit{XMM}-Newton or \hbox{\textit{Swift}/XRT}, respectively. \\
Columns 4--5 : RA and DEC (J2000).\\
Column 6 : Angular separation between the soft \hbox{X-ray} and optical positions. \\
Column 7 : Exposure time.\\
Column 8 : 3--8 keV fluxes in units of $10^{-14}$ erg s$^{-1}$ cm$^{-2}$. Fluxes and $90\%$ upper and lower uncertainties are calculated from \hbox{X-ray} spectral analysis for objects with sufficient photon counts. For the other objects, the fluxes are calculated from count rate using a simple count rate to flux conversion factor because of the limited counts. See \S\ref{subsec:softxray} and Appendix A for details of photometry and flux uncertainty of each object.\\
Column 9 : The energy range of data used. \\
Column 10 : Best-fitting power-law index of the {\tt XSPEC} {\tt pow} model. Some of the objects are fitted with a fixed $\Gamma = 1.8$. \\
Columns 11--12 : Galactic column density and the best-fitting absorption column density for the {\tt pow} component in cm$^{-2}$.\\
Column 13 : We use the unbinned data and the Cash statistic \citep{cash79} for the spectral analysis. The goodness of the fit is assessed using the {\sc XSPEC} {\it goodness} command. See \S~\ref{subsec:softxray} for details.\\
Column 14 : Rest-frame $L_{2-10\mathrm{keV}}$ not corrected for absorption. The absorption-corrected $L_{2-10\mathrm{keV}}$ values are presented in parentheses. Similar to previous columns, $L_{2-10\mathrm{keV}}$ for objects with sufficient photons is derived from the best-fit model. For other objects, $L_{2-10\mathrm{keV}}$ is calculated from the soft \hbox{X-ray} fluxes and converted to rest-frame $L_{2-10\mathrm{keV}}$ assuming a power-law photon index of 1.8, and their absorption-corrected $L_{2-10\mathrm{keV}}$ values are calculated based on the same power-law model hardness-ratio inferred $N_{\rm H}$ values.
\\
$^a$ : Due to the \xmm\ position of J023229+2023.7 being located near the chip gap, we estimate its soft \hbox{X-ray} properties using \nus\ data.\\
$^b$ : $N_{\rm H}$ is estimated based on the hardness ratio assuming a $\Gamma=1.8$ X-ray spectrum using {\sc PIMMS}.\\
$^c$ : Additional soft \hbox{X-ray} observations not used for the analysis due to the large difference between their 3--8 keV fluxes (estimated using their photon count rates) and the 3--8 keV fluxes of their \nus\ counterparts.}
\end{deluxetable*}

\section{Data Analysis}\label{sec:data}
\begin{figure*}
%\figurenum{2}
\vspace*{-1.in}
\hspace*{-0.5in}
\epsscale{1.3}
\plotone{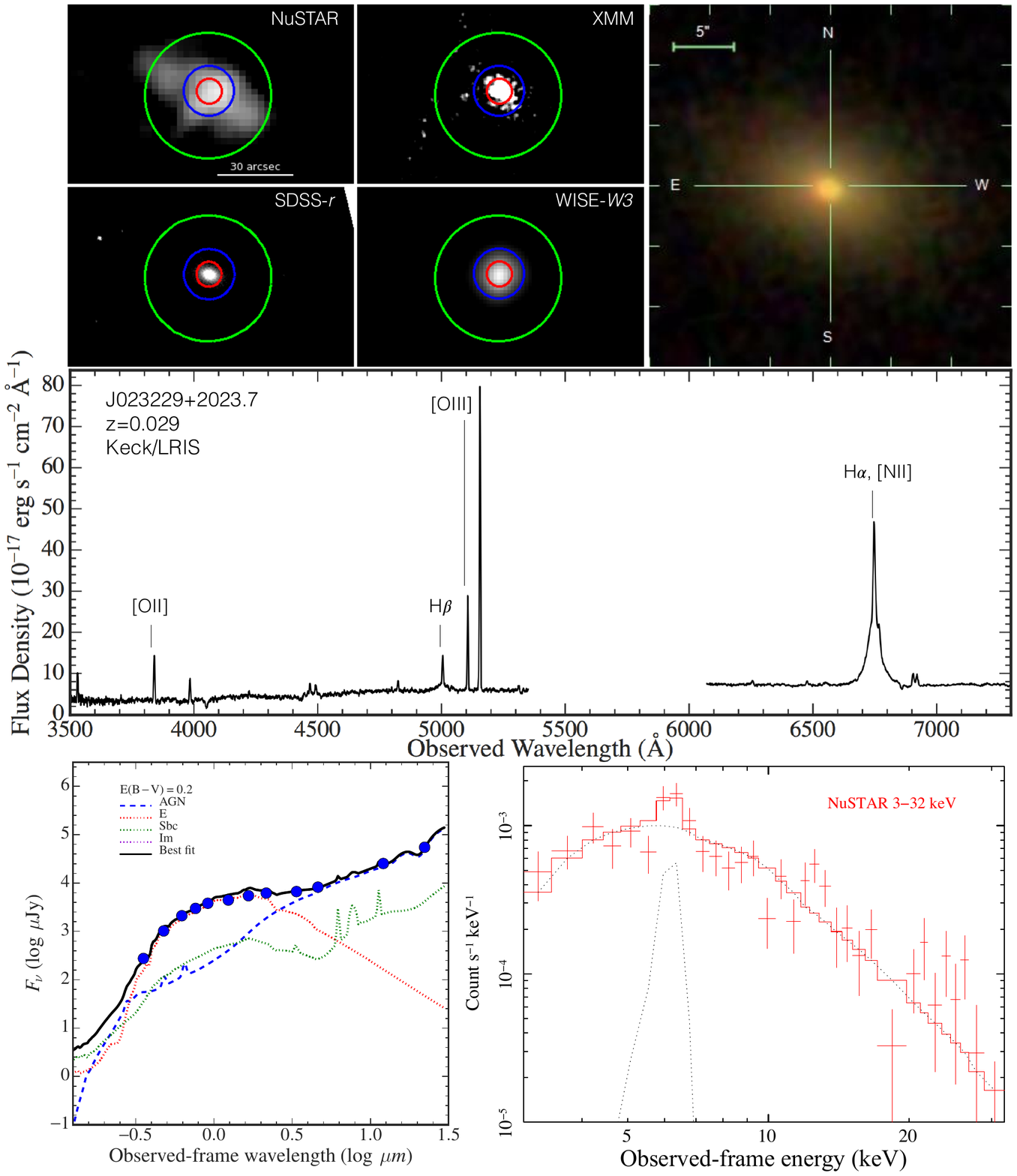}
\vspace*{-1.in}
\caption{Multiwavelength data of J023229+2023.7. The upper-left panel shows the location of this object in {\it NuSTAR}-FB, \hbox{\textit{XMM}-Newton}, SDSS {\it r}, and {\it WISE} W3 bands (upper-left, upper right, lower-left and lower-right, respectively). The green (25\arcsec radius), blue (10\arcsec radius), and red (5\arcsec radius) circles represent the {\it NuSTAR}, soft X-ray, and optical locations, respectively. The upper-right panel shows the false-color image obtained using the SDSS DR 12 image cutout tool centered on the associated optical galaxy. The middle panel shows the observed optical spectrum of J023229+2023.7. The multiwavelength photometry and best-fit SED are shown in the bottom-left panel. The \nus\ \hbox{X-ray} spectrum and the best-fit model are shown in the bottom-right panel. See Appendix A for a brief summary of this object.
The complete figure set (10 images) for the entire {\it NuSTAR} low-mass AGN sample is available in the online journal.}
\label{fig:figset}
\end{figure*}

In this section, we outline data-analysis methods and results for the \nus\ low-mass galaxy sample. 
For each object, we use a spectral energy distribution (SED) fitting analysis to estimate its stellar mass in \S\ref{subsec:sedfitting}. The \nus\ and ancillary soft \hbox{X-ray} data are presented in \S\ref{subsec:hardxray} and \S\ref{subsec:softxray}, respectively. 
The optical spectral analysis results are presented in \S\ref{subsec:optspec}. The data-analysis results for individual objects are shown in the online figure set (Figure~\ref{fig:figset}) and Appendix A.

\subsection{SED fitting and stellar-mass estimation}\label{subsec:sedfitting}
To estimate $M_\star$ for the members of the low-mass AGN sample, we use an SED-fitting based approach similar to the one adopted by \cite{alex13nustar} which takes the possible contribution from the AGN into account. We fit the photometry spanning the UV to mid-IR with the low-resolution templates from \cite{asse10sed}, which are comprised of three galaxy templates and one AGN template.
This approach has been shown to be able to reproduce accurately the SEDs of a wide variety of galaxies \citep[see][for details]{asse08sed} as well as effectively disentangle the AGN contribution from the host-galaxy photometry \citep[e.g.,][]{ster14nustar,chun14,chen15qsosf,lans15}.
For the AGN component, we create a grid of AGN templates with $0<E(B-V)<10$ by applying dust reddening to the \cite{asse10sed} AGN template.
The extinction model we use is a hybrid of an SMC-like (Small Magellanic Cloud) extinction
curve at $\lambda < 3300$ \AA\ (Gordon \& Clayton 1998) and a Galactic extinction curve at longer wavelengths \citep{card89},
with $R_V = 3.1$ for both. 
This is motivated by the observed lack of the 2175\AA\ carbon feature in QSO spectra \citep[see][for details]{york06qso,asse10sed}. 
Once the best-fit SED is determined using the non-negative least square algorithm presented by \cite{asse10sed},
we calculate the rest-frame {\it i}-band absolute magnitude and rest-frame {\it r$-$z} color of the galaxy component of the best-fitting SED,
which is then used to estimate the stellar mass using the color-dependent mass-to-light ratio described
in \cite{zibe09mstar}, with a Chabrier initial-mass function. We present the resulting $M_\star$ values of the \nus\ AGN hosted by low-mass galaxies as a function of redshift in Figure~\ref{fig:rmag_mass}b. For comparison, we have also shown the $M_\star$ values for the parent \nus\ serendipitous sample with spectroscopic redshift $z<0.5$ in the same figure.

For the seven \nus\ low-mass AGN candidates with SDSS photometry,
we also calculate $M_\star$ using the {\sc kcorrect} package\footnote{\url{http://kcorrect.org/}} \citep{blan03kcorr} without correcting for the AGN contribution. For our sample, the SED-fitting estimated $M_\star$ is only slightly lower than the {\sc kcorrect} $M_\star$ by $\approx 0.04$ dex, which is consistent with previous studies that suggest the AGN contamination at optical to near-IR wavelengths
is generally not significant for local low-luminosity broad-line AGNs \citep{rein15dwarf, hain16}.
In this work, we adopt the $M_\star$ values estimated using the SED-fitting method described above for the low-mass AGN candidates in our sample. 
The errors on $M_\star$ are expected to be dominated by $\approx 0.3$ dex uncertainties associated with the stellar population synthesis model degeneracy  \citep{conr09mstar}.
We list the stellar masses in Table~\ref{tab:mwprop}, and the SED fitting results for each object are shown in Figure~\ref{fig:figset}. The stellar mass range for the 10 galaxies in our final sample is 8.9--10.0 ($\log M_\sun$).

\subsection{Basic \nustar\ properties}\label{subsec:hardxray}
Due to the design of the \nustar\ serendipitous survey, the low-mass galaxies in our sample are covered by \nustar\ observations with a wide range of exposure times, between 12~ks and 125~ks (see Table~\ref{tab:nusprop} for the \nus\ properties of our sample). 
Some of them have substantial background counts (see Table~\ref{tab:nusprop}) due to small angular separation from the luminous primary science targets. Some of the sources are located at the edge of the \nus\ FOV thus making spectral extraction challenging. Only four of the sources in our sample have sufficient net photon counts and manageable background to be suitable for \hbox{X-ray} spectral-fitting analyses (J023229+2023.7, J032459--0256.2, J065318+7424.8, and J101609--3329.6).
In the L17 serendipitous catalog, an observed hard \hbox{X-ray} luminosity has been calculated for each object in the rest-frame 10--40 keV band
($L_{10-40\mathrm{keV}}$ hereafter) using the \nus\ hard-band ($8-24$ keV) flux derived from the count rates with a count-rate to flux conversion factor based on a power-law AGN spectrum with a photon index of $\Gamma=1.8$. 
There are two objects with less than $2\sigma$ detection significance in the hard band. We consider their $L_{10-40\mathrm{keV}}$ to be upper limits. For the other objects with {\it NuSTAR}-HB detections, their {\it NuSTAR}-HB flux uncertainties translate into an $L_{10-40\mathrm{keV}}$ average uncertainty of 0.2 dex.
Since the photon counts for our low-mass AGN candidates are limited, we use the band ratio ($H/S$, in which $S$ and $H$ represent the \nustar\ net counts in the $3-8$ keV and  $8-24$ keV bands) to estimate the basic \hbox{X-ray} spectral shape for each source, respectively. 
To account for the uncertainties associated with the high background counts of \nus\ observations,
we calculate the band ratios using the Bayesian Estimation of Hardness Ratios method \citep[BEHR,][]{soft_behr}.
The calculated band ratios for our low-mass AGN candidates span a range of $0.43-1.3$ (see Table~\ref{tab:nusprop}).
We used the {\sc XSPEC fakeit} command to simulate \nus\ spectra and found that this range of band ratios corresponds to spectra with effective power-law photon indices
in the range $0.9<\Gamma<2.5$. With the substantial uncertainties due to the limited photon statistics, limited energy range, and high background,
the \hbox{{\it NuSTAR}} band ratios alone do not allow us to determine whether there is substantial obscuration in any of the low-mass AGNs.
Thus, we utilize the ancillary soft \hbox{X-ray} observations to explore further the \hbox{X-ray} spectral properties of
our low-mass AGNs in the following subsections. The \nus\ band ratios and $L_{10-40\mathrm{keV}}$ values are listed in Table~\ref{tab:nusprop}. For our sample, the $L_{10-40\mathrm{keV}}$ range is 39.8--43.4 (in logarithmic erg s$^{-1}$) and the $3-24$ keV flux range is $(6.2-117.1)\times 10^{-14}$ erg cm$^{-2}$ s$^{-1}$.

\subsection{Ancillary X-ray observations and 2--10 keV X-ray luminosities}\label{subsec:softxray}
We use existing data from other \hbox{X-ray} observatories to help constrain the \hbox{X-ray} spectral properties of the \nustar\ low-mass galaxies.
For the 10 objects in our sample, seven of them have archival {\it XMM-Newton} data.

We use the \emph{Science Analysis Software} ({\tt SAS v.15.0.0}\footnote{\url{http://www.cosmos.esa.int/web/xmm-newton/sas-download}})
to process the \xmm\ Observation/Slew Data Files downloaded from the \xmm\ science archive.\footnote{\url{http://nxsa.esac.esa.int}}
Each observation is processed with the SAS task {\sc epicproc} using the latest calibration files (as of Dec. 2016). 
High energy light curves are generated from the EPIC event files ($10-12$ keV for PN and $>10$ keV for MOS) and then used for screening background flares.
Source spectra are extracted from the background-filtered event files using a circular region with a radius of $\approx 10-20\arcsec$.
Background spectra are extracted using circular source-free regions next to the corresponding source
($\approx$~30$^{\prime\prime}-$60$^{\prime\prime}$ radius regions).
Using the SAS tasks {\tt rmfgen} and {\tt arfgen} we also produced the response matrices for each source for the EPIC pn detector. We note that several objects in our sample fall outside of the FOVs of the EPIC MOS detectors, and thus we only adopt the data from the pn detector for consistency. 
\par

There are also three objects with publicly available {\it Chandra} observations.
Their data were analyzed using {\tt CIAO 4.8}. The data were reprocessed using the {\tt chandra\_repro} pipeline to
create the new level 2 event files. The {\it Chandra} source
spectra were extracted from circular regions with a radius of
$\approx$~2$^{\prime\prime}-$10$^{\prime\prime}$, while the background
spectra were extracted from several source-free regions of
$\approx$~40$^{\prime\prime}$ radius, selected at different positions
around the source to account for local background variations.

For J032459--0256.2 and J120711+3348.5, we obtained the contemporaneous archival {\it Swift}/XRT data and used the HEAsoft (v.6.12) pipeline\footnote{\url{http://heasarc.gsfc.nasa.gov/docs/software/lheasoft/}} {\tt xrtpipeline} for data reduction. This cleans the event files using appropriate calibration files and extracts the spectra and ancillary files for a given source position; the source-extraction regions had radii of $\approx$~20$^{\prime\prime}$.

The data quality varies substantially between different objects, and only seven of the sources in our sample have enough photon counts
for \hbox{X-ray} spectral fitting. For these seven galaxies (J023229+2023.7, J032459--0256.2, J065318+7424.8, J101609--3329.6, J103410+6006.7, J115851+4243.2, and J121358+2936.1), we use {\sc XSPEC 12.9.0}\footnote{\url{https://heasarc.gsfc.nasa.gov/xanadu/xspec/}} to perform spectral analysis of the unbinned data using the Cash statistic \citep{cash79}. 
For the four galaxies with acceptable \nus\ data, we jointly fit the \nus\ and soft X-ray data in our spectral analysis.
For each object, we start the analysis by fitting the data with a basic absorbed power-law model using {\sc tbabs*(ztbabs*zpow)} from {\sc XSPEC}, which takes both intrinsic and Galactic absorption column densities into account. 
We then use the {\sc XSPEC} {\it goodness} command to assess the goodness-of-fit. The {\it goodness} command was set to simulate 1,000 spectra from the best-fit parameters. For each simulated spectrum, {\it goodness} computes the Kolmogorov-Smirnov (K-S) statistic, the ``similarity'' between the model spectra and the data, between the best-fit model and the simulated spectrum. 
The resulting goodness-of-fit is defined as the fraction of simulated spectra with a K-S statistic smaller than that between the best-fit model and data. We consider the fit to be acceptable for sources with {\it goodness} $<50\%$. 
Depending on the goodness-of-fit and inspection of the residuals of the basic model, some of the objects require an additional iron K$\alpha$ line component or an additional diffuse plasma component to achieve an acceptable fit. 

The \hbox{X-ray} spectral analysis procedures for each object are presented in Appendix A. For the objects with sufficient \hbox{X-ray} photon counts, we calculate the following \hbox{X-ray} spectral properties using the best-fit model: intrinsic $N_{\rm H}$ value, \hbox{X-ray} luminosity (not corrected for absorption) measured in the rest-frame 2--10 keV band ($L_{2-10\mathrm{keV}}$ hereafter), \hbox{X-ray} flux measured in the observed-frame 3--8~keV band, and \hbox{X-ray} power-law photon index $\Gamma$. Uncertainties of fluxes and model parameters are estimated using the $90\%$ confidence intervals.
Three of these seven objects have $N_{\rm H}>10^{22}$ cm$^{-2}$.

For the other three objects in our sample, J120711+3348.5, J134934-3025.5, and J223654+3423.5, there are not enough \hbox{X-ray} counts for spectral fitting. We estimate their $3-8$~keV \hbox{X-ray} fluxes and $L_{2-10\mathrm{keV}}$ values from their count rates in the corresponding energy ranges using {\tt PIMMS} with a photon index of $\Gamma=1.8$ and a Galactic absorption column density. The flux uncertainties for these three objects were obtained using uncertainties of the photon counts estimated with the \cite{gehr86} method. 
To constrain the basic X-ray spectral properties of these three objects, we estimate their hardness ratios, $(H-S)/(H+S)$, where $H$ and $S$ represent hard-band and soft-band counts, respectively. 
For the \swiftxrt\ data of J120711+3348.5, we measure its $S$ and $H$ at \hbox{0.3--2}~keV and \hbox{2--10}~keV, respectively. 
For the \xmm\ data of J134934-3025.5, we measure its $S$ and $H$ at \hbox{0.5--2}~keV and \hbox{2--10}~keV, respectively. For the \chandra\ data of J223654+3423.5, we measure its $S$ and $H$ at 0.5--2 and 2--8 keV, respectively. 
The hardness ratios and the associated uncertainties are then estimated using BEHR \citep{soft_behr}. 
We also use {\sc PIMMS} to calculate the corresponding $N_{\rm H}$ values using the hardness ratios assuming a $\Gamma=1.8$ power-law spectrum. 

We list the details of ancillary \hbox{X-ray} observations, \hbox{X-ray} spectral fitting parameters, 
and the rest-frame $L_{2-10\mathrm{keV}}$ (observed and absorption-corrected) in Table~\ref{tab:sxprop}. 
The angular offsets between the optical positions and the soft X-ray positions are also listed for reference. 
For objects for which we can do spectral analysis, the rest-frame $L_{10-40\mathrm{keV}}$ values (both observed and absorption-corrected) are also calculated based on the best-fit models and are present in Table~\ref{tab:nusprop}. 
Notably, six of the low-mass AGNs (J023229+2023.7, J032459--0256.2, J065318+7424.8, J101609--3329.6, J115851+4243.2, and J223654+3423.5) in our sample have multiple ancillary \hbox{X-ray} observations. However, these additional observations have limited spatial resolution and/or small photon counts which prevent us from further assessing their X-ray properties. The details of these additional observations are presented in Appendix B and the second half of Table~\ref{tab:sxprop}.

\subsection{Optical spectroscopic observations and analysis}\label{subsec:optspec}

\begin{figure}
\hspace*{-.4in}
\epsscale{1.3}
\plotone{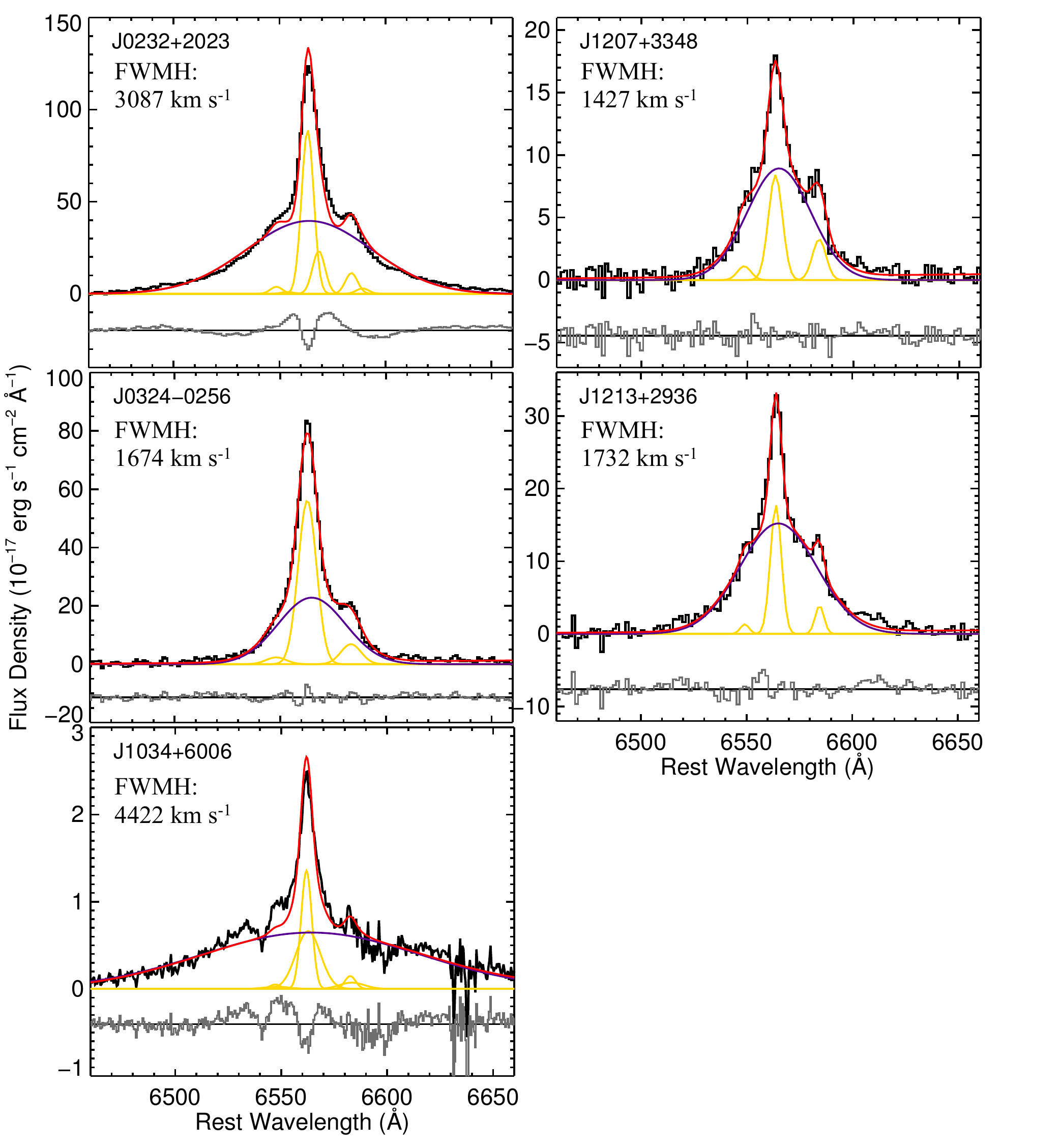}
\caption{The continuum and absorption-line subtracted spectra showing the rest-frame broad H$\alpha$ lines. For each object, the best-fitting model is shown in red. Individual Gaussian components of narrow lines are shown in yellow. The broad H$\alpha$ component is plotted in dark blue. Residuals are plotted in gray with a vertical offset for clarity.}
\label{fig:halpha}
\end{figure}

\begin{figure}
\hspace*{-0.5in}
\epsscale{1.4}
\plotone{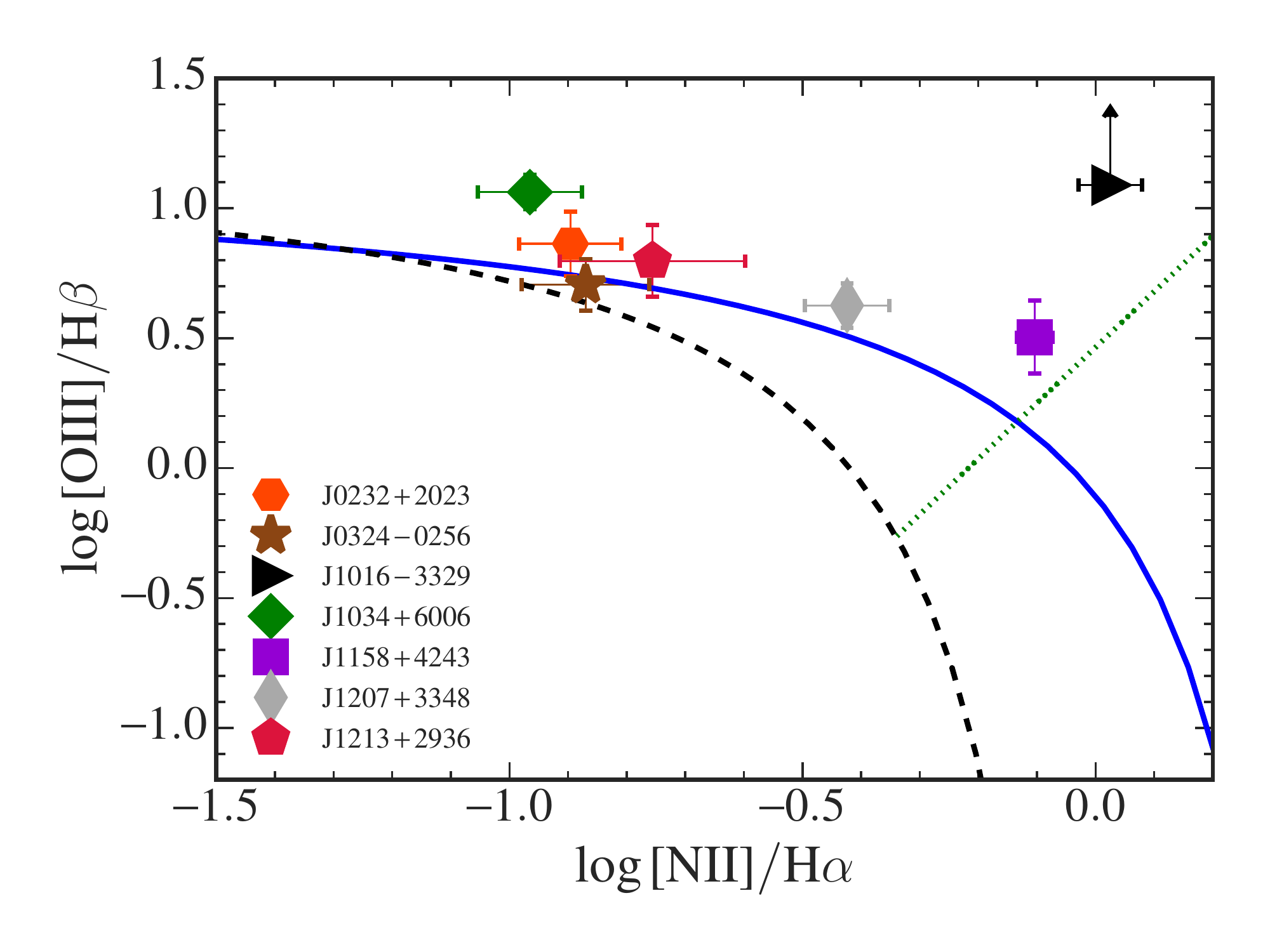}
\caption{The emission-line ratios (BPT diagram) for the \nus\ low-mass AGNs with high SNR optical spectra. All lie above the \cite{kauf03host} empirical curve (dashed curve that defines the ``star-forming region'' and the dotted line that separates LINERs from AGNs) for selecting galaxies with significant AGN contributions to the emission lines. The \cite{kewl06agn} ``maximum starburst'' line (solid curve) is also shown.}
\label{fig:bpt}
\end{figure}

Of the 10 low-mass galaxies, three have existing SDSS spectra. For the other seven galaxies, optical spectroscopic follow-up observations were obtained using either Keck LRIS, Palomar DBSP, or NTT EFOSC2 as part of the spectroscopic follow-up program for the \nus\ serendipitous survey (L17). Details of the spectroscopic observations of our sample are listed in Table~\ref{tab:mwprop}, and the optical spectrum for each object is presented as part of the figure set in Figure~\ref{fig:figset}.

In our sample, only seven of the 10 objects have high SNR spectra with strong emission lines. We analyze the optical spectra of these seven objects with customized software similar to that used in \cite{rein13dwarf}, which removes the host-galaxy contribution and models the AGN emission-line profiles. 
Important features in the optical spectrum that indicate the presence of an accreting mBH are H$\alpha$ and H$\beta$ emission lines that are significantly broadened since they arise from dense gas orbiting an mBH, though these broad Balmer lines are often blended with other narrow emission lines at similar wavelengths. Our analysis method models the blended broad Balmer lines and other high-ionization narrow line components, yielding accurate emission-line width and flux measurements for the broad Balmer lines and other narrow high-ionization lines. The details of the model-fitting method are described in \S3 of \cite{rein13dwarf}. We find that five of the seven objects with high SNR spectra have a robust broad H$\alpha$ component with a line width broader than $500$ km s$^{-1}$ that is indicative of accretion onto an mBH \citep{rein13dwarf}. We show the results of our broad H$\alpha$ emission-line measurements in Figure~\ref{fig:halpha}. For completeness, we also show the close-up spectra for sources without broad H$\alpha$ lines in the Appendix.

We present the distribution of narrow emission-line ratios of [\ion{O}{3}] $5007$/${\rm H}\beta$ versus [\ion{N}{2}] $6583$/${\rm H}\alpha$ \citep[i.e., the ``BPT diagram'',][]{diagnostic_bpt} in Figure~\ref{fig:bpt}, as a diagnostic of the nature of the optical spectra. 
We find that all seven low-mass AGNs with high-quality optical spectra have emission-line ratios above the \cite{kauf03host} empirical curve separating AGNs from star-forming galaxies and LINERs in the BPT diagram, 
and six of these seven AGNs also have emission-line ratios above the \cite{kewl06agn} ``maximum starburst'' curve. This indicates that the emission lines in these objects are powered by accretion onto massive BHs.

\section{The AGN nature of the low-mass galaxies with \textit{N\lowercase{u}STAR} detections}\label{sec:agnprop}

\begin{deluxetable}{lcccc}
\tabletypesize{\scriptsize}
\tablecolumns{5}
\tablewidth{0pt}
\tablecaption{AGN identification criteria}\label{tab:agndiag}
\tablehead{
\colhead{Source Name} &
\colhead{$L_{2-10\mathrm{keV}}$} &
\colhead{Broad H$\alpha$} &
\colhead{BPT diagram} &
\colhead{{\it WISE} color}}
\colnumbers
\decimals
\startdata
J023229+2023.7 & Yes      & Yes         &  Yes     & Yes \\
J032459--0256.2  & No       & Yes         &  Yes$^a$ & No \\
J065318+7424.8  & Yes      & No          &  No      & Yes \\
J101609--3329.6  & Yes      & No          &  Yes     & No \\
J103410+6006.7  & Yes      & Yes         &  Yes       & No  \\
J115851+4243.2  & No       & No          &  Yes       & No \\
J120711+3348.5  & Yes      & Yes         &  Yes       & Yes \\
J121358+2936.1  & Yes      & Yes         &  Yes       & Yes \\
J134934--3025.5  & Yes      & No          &  No        & No \\
J223654+3423.5 & Yes$^b$ & No          &  No        & No \\
\enddata
\tablecomments{The summary table for the results from \S\ref{subsec:veragn}. 
Column 1: Source name. 
Column 2: The $L_{2-10\mathrm{keV}}>10^{42}$ erg s$^{-1}$ criterion.
Column 3: The presence of a broad H$\alpha$ line, see Figure~\ref{fig:halpha}.
Column 4: The BPT emssion-line ratio diagnostics, see Figure~\ref{fig:bpt}.
Column 5: The \cite{ster12wise} {\it WISE} color selection criterion, $W1-W2>0.8$. See \S\ref{subsec:wisecolor} for details. \\
$^a$ : The emission-line ratios of this object do not exceed the ``maximum starburst curve'' defined by \cite{kewl06agn}, but still reside in the AGN region defined by the \cite{kauf03host} curves on the BPT diagram. \\
$^b$ : The $L_{2-10\mathrm{keV}}$ derived based on the \chandra\ observations does not exceed $10^{42}$ erg s$^{-1}$, but the {\it NuSTAR}-SB flux does correspond to a soft \hbox{X-ray} luminosity satisfying $L_{2-10\mathrm{keV}}>10^{42}$ erg s$^{-1}$.}
\end{deluxetable}

\subsection{The optical and X-ray AGN diagnostics}\label{subsec:veragn}

One of the challenges in confirming the presence of an AGN in low-mass galaxies is that the AGN emission is often diluted by stellar processes. 
This is particularly true for objects that are less-luminous in the X-ray band (i.e. $L_{2-10\mathrm{keV}} < 10^{42}$ erg s$^{-1}$) and objects without the telltale high-excitation emission lines in the optical spectra. 

To verify that the 10 {\it NuSTAR}-selected low-mass galaxies are indeed powered by accretion onto an mBH, 
we consider the following diagnostics: 
\begin{enumerate}
    \item The empirical $L_{2-10\mathrm{keV}} > 10^{42}$ \hbox{erg s$^{-1}$} criterion, which generally distinguishes \hbox{X-ray} emission powered by AGNs from that powered by \hbox{X-ray} binaries (see \S\ref{subsec:softxray} and Appendix A for the details of derivation of $L_{2-10\mathrm{keV}}$ for each object). 
    \item The optical emission-line ratio diagnostics (the BPT diagram, Figure~\ref{fig:bpt}). We consider the \cite{kauf03host} and \cite{kewl06agn} curves that separate AGNs from star-forming galaxies and LINERs.
    \item The presence of significantly broadened H$\alpha$ line emission (FWHM > 500 km s$^{-1}$), which indicates the presence of dense gas being accreted onto an mBH (see Figure~\ref{fig:halpha}). 
\end{enumerate}

For the 10 galaxies in our sample, 8 of them have $L_{2-10\mathrm{keV}}>10^{42}$ erg s$^{-1}$ that can be securely attributed to an \hbox{X-ray} AGN. 
As for the optical emission-line diagnostics, all seven objects in our sample with high SNR optical spectra reside in the AGN region defined using the \cite{kauf03host} curves. 
Of these seven objects, all except one, J032459--0256.2, are also above the \cite{kewl06agn} ``maximum starburst'' curve. Although the emission-line ratios of J032459--0256.2 do not exceed the \cite{kewl06agn} curve, its optical spectrum shows a significant broad H$\alpha$ component (FWHM $=1674$ km s$^{-1}$) that is likely to be powered by accretion onto an mBH. Therefore, we consider \hbox{J032459--0256.2} to be an optical AGN as well. \par

We list the results of these diagnostics in Table 5. While we do not use the mid-IR colors to determine the presence of an AGN for the low-mass galaxies in our sample, we also list whether our sample objects satisfy the \cite{ster12wise} {\it WISE} color-selection criterion, $W1-W2>0.8$, in Table 5. A detailed discussion of the validity of using {\it WISE} color to identify AGNs in low-mass galaxies is given in \S\ref{subsec:wisecolor}.

For our sample, there are a total of five objects that satisfy both the \hbox{X-ray} luminosity and optical emission-line diagnostics; we consider these seven objects to be  ``multiwavelength'' AGNs (see the previous subsection and Table~5). For the other five objects in our sample, three of them (J065318+7424.8, J134934--3025.5, and J223654+3423.5) have $L_{2-10\mathrm{keV}}>10^{42}$ erg s$^{-1}$ in at least one epoch of \hbox{X-ray} observations, but have no apparent optical emission lines that could be used for emission-line diagnostics. 
These ``optically-dull'' \hbox{X-ray} AGNs are discussed in more detail in \S\ref{subsec:xbong}. \par

Besides the five multiwavelength AGNs and the three optically-dull \hbox{X-ray} AGNs, there are two X-ray faint sources in our sample that are classified as an AGN at optical wavelengths: J032459--0256.2 and J115851+4243.2. Their X-ray luminosities are fainter than the empirical $10^{42}$~erg s$^{-1}$ threshold for identifying typical X-ray AGNs. For J032459–0256.2, the soft X-ray luminosity, $L_{2-10\mathrm{keV}} = 3.3\times 10^{41}$ erg s$^{-1}$, is derived by jointly fitting the available \nus\ and \swiftxrt\ data. 
We find that the \hbox{X-ray} spectrum of J032459--0256.2 is consistent with an absorbed power-law with $\Gamma = 1.9^{+0.3}_{-0.2}$ and intrinsic $N_{\rm H} < 3.4\times10^{21}$ cm$^{-2}$. Although its 2--10 keV \hbox{X-ray} luminosity does not exceed the $10^{42}$ erg s$^{-1}$ threshold, the optical spectrum exhibits strong broad and narrow high-ionization emission lines powered by an AGN. The BH mass of this object is $\log M_\bullet/M_\sun = 6.06$ based on its broad H$_\alpha$ emission. With the clear optical AGN signatures and the low $M_\bullet$, we consider the X-ray emission from J032459--0256.2 to be indeed powered by accretion onto mBH and J032459--0256.2 is also a ``multiwavelength'' AGN.

For J115851+4243.2, the soft X-ray luminosity is $L_{2-10\mathrm{keV}} = 2.7\times 10^{38}$ erg s$^{-1}$. 
In this luminosity range, it is also possible for the \hbox{X-ray} emission to be powered by \hbox{X-ray} binaries or even a single ULX. However, the optical emission-line ratios of J115851+4243.2 suggest the presence of an underlying AGN. 
The high spatial resolution \chandra\ image also reveals that the position of $>2$ keV \hbox{X-ray} emission coincides with the SDSS fiber location with $< 0.3\arcsec$ separation (see Table~\ref{tab:mwprop}). 
Considering the absolute astrometric uncertainties of \chandra\ and SDSS ($< 1.1\arcsec$ and $<0.1\arcsec$, respectively), the physical separation between the hard X-ray source and the SDSS fiber location is less than $\approx 50$ pc. In \S\ref{subsec:obscuration} we explore further the nature of J115851+4243.2 and show that the weak X-ray emission of J115851+4243.2 might be due to the presence of large amounts of obscuring material.

\subsection{Optically-dull X-ray AGNs}\label{subsec:xbong}
In our sample, there are three objects with luminous X-ray emission but no obvious AGN-powered emission lines. 
In detail, the optical spectrum of J134934--3025.5 is consistent with that of a quiescent galaxy with weak emission lines (see online figure set 2.9). For J223654+3423.5, while its optical spectrum shows strong H$\alpha$ and [\ion{N}{2}] emission lines, there is no apparent [\ion{O}{3}] emission. For J065318+7424.8, the optical spectrum has limited SNR but no immediately visible evidence indicative of AGN-powered high-ionization lines (see online figure set 2.3).

There have been extensive studies of the population of objects with bright \hbox{X-ray} nuclei and weak or no optical emission lines \citep[e.g.,][]{elvi81xbong,coma02,rigb05,civa07,trum11}. 
The lack of AGN-powered optical emission lines in these \hbox{X-ray} bright objects has often been attributed to a combination of the following reasons: i) host-galaxy dilution due to the large optical fiber/slit radius \citep{mora02}, ii) parsec-scale obscuration, iii) a scenario in which the SMBH is accreting at a low Eddington rate via a radiatively inefficient accretion flow \citep[RIAF, e.g.][]{yuan04}, or iv) the radiation from the recently triggered SMBH accretion has yet to reach the narrow emission line regions \citep[e.g.][]{scha15}. 

Considering the luminous \hbox{X-ray} emission and low stellar mass of the objects described in the previous paragraph, it is unlikely that their ``optical dullness'' is due to underlying RIAF-powered AGNs. 
For instance, the Eddington ratio for J134934--3025.5, the most massive object in our \nus\ sample, would be $L/L_{\rm Eddington}\sim 0.2$ if we derive $M_\bullet$ using the \cite{rein15dwarf} $M_\bullet-M_\star$ relation for local AGNs. This is much higher than the threshold for an RIAF, $L_{\rm bolometric}/L_{\rm Eddington}\lesssim 0.01$ \citep[e.g.,][]{bege84,nara95nat,yuan04}. 
On the other hand, the \nus\ band ratios of the three objects without apparent AGN-powered emission lines are $\gtrsim 0.9$. These values are consistent with flat power-law \hbox{X-ray} spectra with $\Gamma\sim 1.2$, suggesting the AGNs in these objects might be obscured. However, the uncertainty of the \nus\ band ratios is substantial for the low-luminosity objects in our sample due to high background counts caused by the angular proximity to the bright primary science targets. The current soft \hbox{X-ray} data for these three objects are also limited by small numbers of photons. Therefore, additional \hbox{X-ray} and optical spectroscopic observations are required to determine the cause of the lack of optical emission lines in these three objects.

\subsection{Identifying heavily obscured AGNs in low-mass galaxies with \textit{N\lowercase{u}STAR}}\label{subsec:obscuration}

\begin{figure}
\hspace*{-0.25in}
\epsscale{1.3}
\plotone{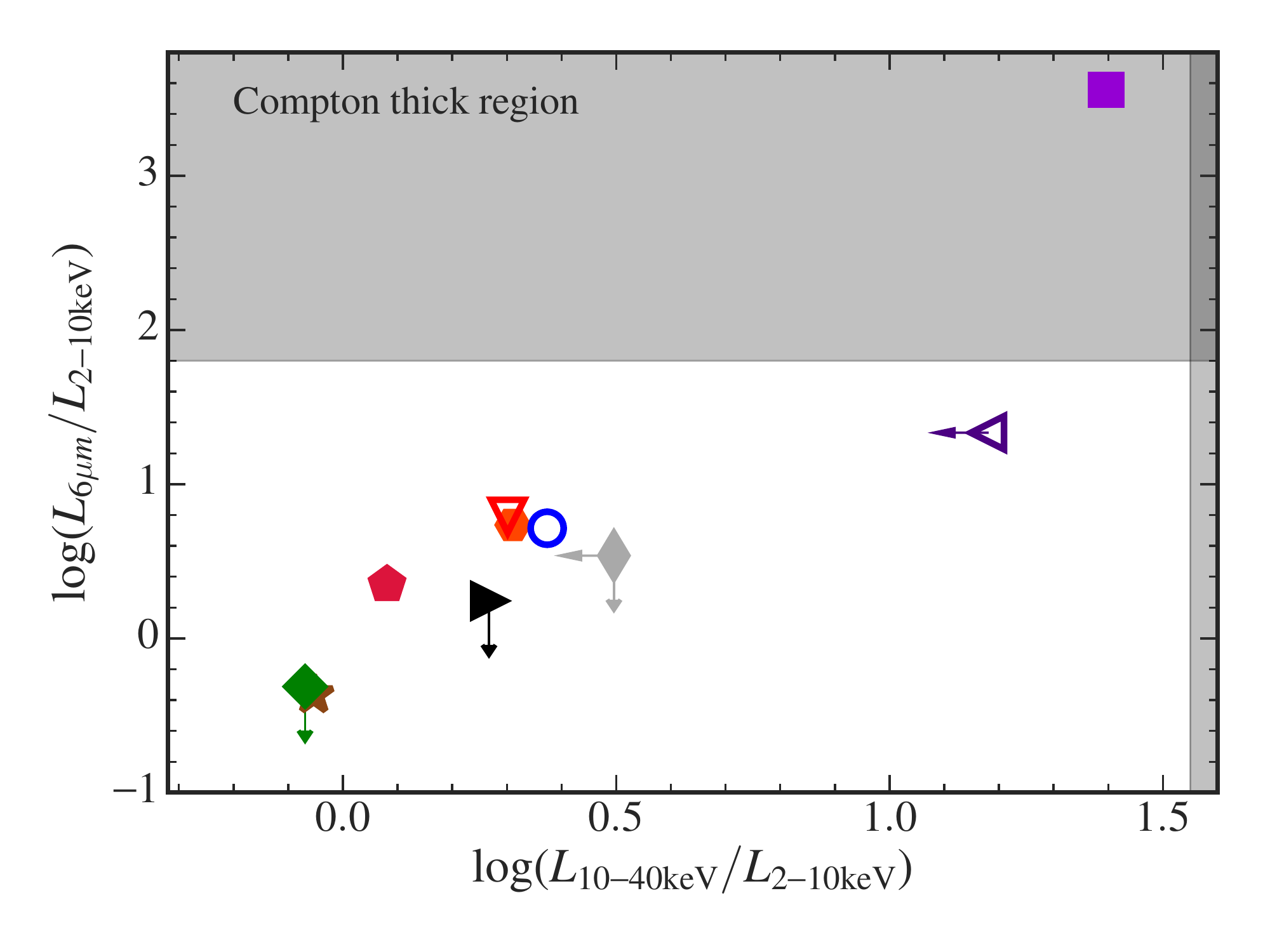}
\caption{The distributions of two AGN luminosity ratios, $L_{\rm MIR}/L_{2-10\mathrm{keV}}$ versus $L_{10-40\mathrm{keV}}/L_{2-10\mathrm{keV}}$. The shaded area is the ``Compton thick region'', which is defined by assuming an absorbed power-law X-ray spectrum with $\Gamma=1.8$ and $N_{\rm H} > 1.5\times10^{24}$ cm$^{-2}$. The more heavily obscured objects occupy the upper-right corner of the plot. In particular, J115851+4243.2 (filled purple square) appears to be obscured by Compton-thick column densities (see \S\ref{subsec:obscuration}). Symbols are as in Figure~\ref{fig:rmag_mass}.}
\label{fig:lagn_ratios}
\end{figure}

The ratio between observed $L_{10-40\mathrm{keV}}$ and $L_{2-10\mathrm{keV}}$ has been considered as a basic indicator of obscuration since hard X-rays are relatively insensitive to the presence of obscuring material than the soft X-rays. For the low-mass AGNs in our sample, we study their $L_{10-40\mathrm{keV}}$/$L_{2-10\mathrm{keV}}$ ratios to investigate how many of them are heavily obscured. 

While $L_{10-40\mathrm{keV}}$ is much less affected by absorption than $L_{2-10\mathrm{keV}}$, it can still be significantly suppressed when the obscuring column reaches $N_{\rm H}\gtrsim 10^{24}$ cm$^{-2}$ \citep[e.g.,][]{ster14nustar,lans14nustar,lans15,lama16,lans17ser}.
Therefore, we also utilize the mid-IR luminosities measured from SED-decomposition for our sample. 
AGN mid-IR and soft \hbox{X-ray} (e.g., rest-frame 2--10 keV) emission have been found to have an almost 1-to-1 correlation in unobscured Seyfert AGNs \citep[e.g.,][]{fior09obsc,gand09seyfir,asmu15,mate15,ster15,chen17lxl6um} similar to $L_{10-40\mathrm{keV}}$ and $L_{2-10\mathrm{keV}}$. Therefore, the ratio between $L_\mathrm{MIR}$ and $L_{2-10\mathrm{keV}}$ has been widely adopted as an indicator of obscuration toward the nucleus \citep{alex08compthick,delm16}.

To identify objects with extreme obscuration, we investigate the distributions of the two different luminosity ratios mentioned above.
We calculate the $6\micron$ monochromatic luminosity ($\nu L_\nu$ measured at rest-frame $6\micron$, $L_{6\micron}$ hereafter)
for the AGN component of the best-fit SEDs described in \S\ref{subsec:sedfitting}.
The two luminosity ratios, $L_{10-40\mathrm{keV}}/L_{2-10\mathrm{keV}}$ and $L_{6\micron}/L_{2-10\mathrm{keV}}$, are shown in Figure~\ref{fig:lagn_ratios}. 

Overall, $L_{10-40\mathrm{keV}}/L_{2-10\mathrm{keV}}$ and $L_{6\micron}/L_{2-10\mathrm{keV}}$ appear plausibly to follow a simple relation for our {\it NuSTAR}-selected low-mass AGNs,
including objects without distinct AGN-powered emission lines and objects that are heavily obscured at soft \hbox{X-ray} energies.
This is not surprising, as objects that are more heavily obscured have a weaker $L_{2-10\mathrm{keV}}$ relative to $L_{10-40\mathrm{keV}}$ and $L_{6\micron}$. 
The median difference between the $L_{\rm MIR}$ of our sample and that derived from their intrinsic $L_{2-10\mathrm{keV}}$ using the \cite{chen17lxl6um} $L_{\rm MIR}/L_{2-10\mathrm{keV}}$ relations is $\sim 0.4$ dex, suggesting a non-neglegible residual mid-IR host-galaxy contamination due to the large PSF size of {\it WISE}.
We also caution that the two luminosity ratios shown in Figure~\ref{fig:lagn_ratios} have the same denominator ($L_{2-10\mathrm{keV}}$), but the purpose of Figure~\ref{fig:lagn_ratios} is not to study the correlation between the intrinsic AGN luminosity ratios, but rather just to identify objects in our sample that might have extreme obscuration. 
For instance, in the upper-right part of Figure~\ref{fig:lagn_ratios},
there are two objects, J115851+4243.2 and J223654+3423.5, with $L_{6\micron}/L_{2-10\mathrm{keV}} > 10$. 

For J223654+3423.5, there are not enough \hbox{X-ray} photons to determine the obscuring column density using \hbox{X-ray} spectral modeling. However, the \chandra\ hardness ratio of J223654+3423.5 is $\approx 0.1$, which is consistent 
with that of obscured AGN, $N_{\rm H}\approx 10^{22}$ cm$^{-2}$ (see \S\ref{subsec:softxray} and Table~\ref{tab:sxprop}). 

For J115851+4243.2, the mid-IR luminosity is higher than $L_{2-10\mathrm{keV}}$ by 3.4 dex, and the difference between $L_{10-40\mathrm{keV}}$ and $L_{2-10\mathrm{keV}}$ is only 1.4 dex. While the large intrinsic column density derived from fitting the \chandra\ data ($N_{\rm H}\approx 1.2^{+1.4}_{-1.0}\times10^{23}$ cm$^{-2}$, see Appendix A)
supports the presence of heavy obscuring material, the 3.4 dex difference between $L_{2-10\mathrm{keV}}$ and $L_{6\micron}$
requires that J115851+4243.2 be obscured by Compton-thick material if the intrinsic $L_{2-10\mathrm{keV}}$ follows the linear $L_\mathrm{MIR}$ and $L_{2-10\mathrm{keV}}$ relations for Seyfert 1 AGNs. 
Notably, the optical SED of J115851+4243.2 is entirely dominated by the host-galaxy component (see online figure set 2.6), which also suggests the AGN is heavily obscured. 
Furthermore, J115851+4243.2 is extended in all four bands of the {\it WISE} images.
Therefore, the mid-IR luminosity derived based on the {\it WISE} photometry may have significant host-galaxy contamination.
Indeed, the mounting evidence for the tight correlation between AGN mid-IR luminosity and \hbox{X-ray} luminosity for local Seyfert galaxies is derived based on high angular resolution $\approx12\micron$ observations \citep[e.g.,][]{gand09seyfir,asmu15}. In these studies, the nuclear mid-IR luminosity and $L_{2-10\mathrm{keV}}$ are found to have an almost linear correlation.
We emphasize that our SED-fitting approach has taken the contribution from host galaxies into account when measuring $L_{6\micron}$. For J115851+4243.2, the best-fit SED does imply that the stellar emission remains non-negligible in the \textit{W1} and \textit{W2} bands (see online figure set 2.6), but the host-galaxy contribution rapidly drops at longer wavelengths. \par

\begin{figure}
\plotone{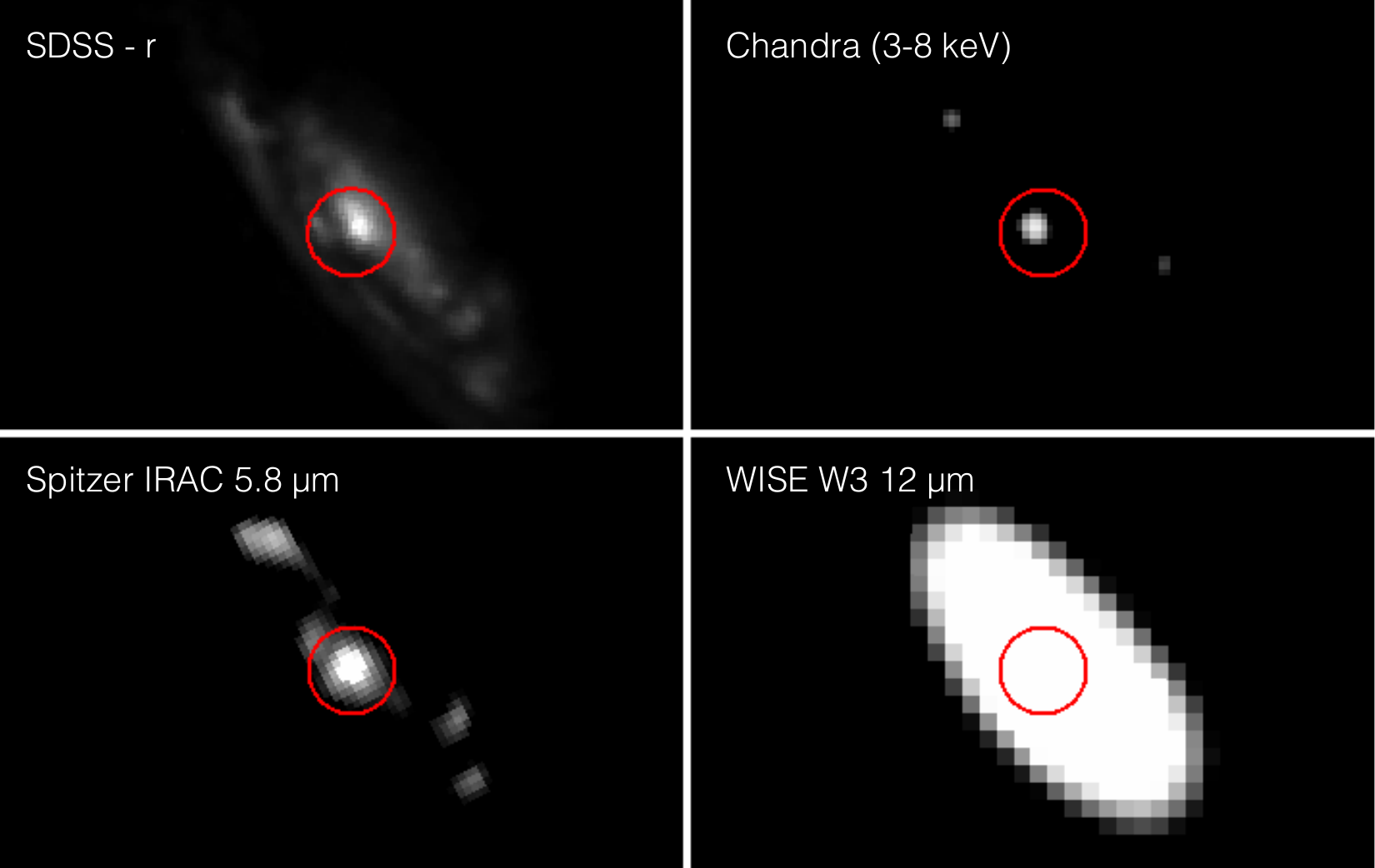}
\caption{Images of J115851+4243.2 at SDSS {\it r} (top-left), {\it Chandra} 3--8 keV (top-right), {\it Spitzer} IRAC $5.8\micron$ (bottom-left), and {\it WISE} W3 (bottom-right) bands. The red circle with a $5\arcsec$ radius defines the region use for aperture photometry at the {\it Spitzer} IRAC $5.8\micron$ band. This figure demonstrates that J115851+4243.2 has a powerful nuclei at both mid-IR and 3--8 keV bands, suggesting the presence of an obscured AGN.}
\label{fig:ic750}
\end{figure}

To test whether the SED-decomposed mid-IR AGN luminosity for J115851+4243.2 still suffers from host-galaxy contamination,
we obtained the {\it Spitzer} Infrared Array Camera \citep[IRAC,][]{inst_irac} image for J115851+4243.2 at $5.8\micron$ from the {\it Spitzer} Heritage Archive\footnote{\url{http://irsa.ipac.caltech.edu/applications/Spitzer/SHA/}}. The PSF of the IRAC $5.8\micron$ band has a 1.88\arcsec\ FWHM, which is smaller than the $> 6\arcsec$ FWHM of {\it WISE} PSFs. 
For comparison, we show the images of J115851+4243.2 at the SDSS {\it r}-band, {\it Chandra} 3--8 keV band, {\it Spitzer} IRAC $5.8\micron$ band, and {\it WISE} W3 band in Figure~\ref{fig:ic750}. 
In the IRAC $5.8\micron$ image, J115851+4243.2 also appears to host a powerful point source near its center. 
We measure the IRAC $5.8\micron$ flux for the central source using an aperture of $5\arcsec$ radius with 
MOsaicker and Point source EXtractor \citep[{\sc MOPEX},][]{soft_mopex}.\footnote{\url{http://irsa.ipac.caltech.edu/data/SPITZER/docs/dataanalysistools/tools/mopex/}}
The measured $5.8\micron$ flux is $(9.8 \pm 1.0)\times10^{-2}$ Jy, which corresponds to a mid-IR luminosity of $L_{6\micron} = 6.0\times10^{41}$ erg s$^{-1}$. 
This is only $\sim 0.2$ dex lower than the mid-IR luminosity measured based on SED-fitting with {\it WISE} photometry, $L_{6\micron} = 9.7\times10^{41}$ erg s$^{-1}$, suggesting the large ($\approx 3.4$ dex) difference between 
the mid-IR and \hbox{X-ray} luminosities of J115851+4243.2 shown in Figure~\ref{fig:lagn_ratios} might indeed be caused by nuclear obscuration. 

Since further resolving the nuclear mid-IR emission of J115851+4243.2 is not possible even with the {\it Spitzer} IRAC images, it is informative to consider its optical spectrum (Figure set 2.6). The strong \ion{Ca}{2} and Na D absorption lines in the optical spectrum and the red optical color ($u-r \approx 3.1$) imply that the galaxy is dominated by an old stellar population.
The emission-line ratios are also consistent with those of an optical AGN (see \S\ref{subsec:optspec}).
Thus, it is not likely that J115851+4243.2 is a starburst galaxy powerful enough
to generate $L_{6\micron}\approx10^{42}$ erg s$^{-1}$  without an underlying AGN.
In fact, if we assume all the mid-IR emission at {\it W3} is powered by star-formation activity,
its specific star formation rate estimated using its $8-1000\micron$ infrared luminosity
($L_{\rm IR} \sim 2.8\times 10^9L_\sun $, estimated using the Chary \& Elbaz 2001 star-forming galaxy templates) and $M_\star$ would be more than 10 times higher than that of the Milky Way.
This is not plausible for a galaxy with a spectrum dominated by an old stellar population.
Therefore, the luminous mid-IR emission is more likely to be dominated by an obscured AGN.

It is also interesting that J115851+4243.2 is tentatively identified as a water megamaser AGN \citep{darl14}.
Several previous studies have pointed out that AGNs with megamaser emission are likely
to be obscured by Compton-thick column densities because the observation of masers requires an edge-on view of the accretion disk \citep{zhan06maser,gree08maser,masi16}.
Therefore, the large difference between the $L_{6\micron}$ and $L_{2-10\mathrm{keV}}$ of J115851+4243.2 appears to be due to
the presence of a Compton-thick obscuring column density.
The science target with a 39 ks \nus\ observation, IC751 \citep{ricc16ic751}, is 9.13\arcmin\ away from J115851+4243.2.
With the rapid degradation of higher energy band sensitivity at large off-axis angles,
J115851+4243.2 has only been detected at $\approx 2\sigma$ significance in the $8-24$ keV band.
For the \chandra\ data, the best-fit intrinsic column density is $N_{\rm H}=1.2^{+1.4}_{-1.0}\times10^{23}$ cm$^{-2}$ (see Appendix A), but the $>3$ keV photon counts are very limited ($\approx 10$ photons only) and therefore cannot rule out higher obscuring column densities, especially considering that the real \hbox{X-ray} spectrum is likely more complex than our basic modeling \citep[e.g.,][]{balo15}. Therefore, a future on-axis observation with a longer exposure time by \nus\ is required to reveal if J115851+4243.2 is indeed a Compton-thick  AGN hosted by a low-mass galaxy. Also, mid-IR imaging with sub-arcsec resolution can provide insight into the origin of the luminous mid-IR emission of J115851+4243.2.

\section{Comparison with previous studies}\label{sec:compare}
To evaluate the effectiveness of using \nus\ to select low-mass AGNs,
we compare the properties of our sample with the other relevant AGN samples reported in the literature. 
We discuss mid-IR colors in \S\ref{subsec:wisecolor}, the $M_\star-M_\bullet$ relation in \S\ref{subsec:mbh}, and \hbox{X-ray} properties in \S\ref{subsec:xprop}.

\subsection{Mid-IR colors}\label{subsec:wisecolor}
Similar to hard X-rays, mid-IR observations are a powerful tool for studying AGN activity that could be enshrouded by intervening dust.
Many studies have utilized {\it Spitzer} and {\it WISE} observations to show that the distinctive red mid-IR color arising from hot dust heated by SMBH accretion can be used as an effective indicator of intrinsically luminous AGN activity \citep[e.g.,][]{lacy04,ster05,ster12wise,donl12,asse13wise,mate13}.
However, AGN identification methods based on red mid-IR colors become more ambiguous for low-luminosity AGNs \citep[e.g.,][]{hain16}.
In particular, recent studies have found that the vast majority of low-mass galaxies with red mid-IR color do not show any sign of optical emission lines powered by AGNs \citep{saty14,sart15,secr15,saty16}. 
Several studies have also demonstrated that young, compact starbursts in dwarf galaxies can have mid-IR colors mimicking those of luminous AGNs \citep[e.g.,][]{grif11,izot14,hain16}.
Here we explore if our hard \hbox{X-ray} selected low-mass galaxies have mid-IR emission powered by AGN-heated dust to assess the effectiveness of using mid-IR emission to search for obscured AGNs in low-mass galaxies.

\begin{figure}
\hspace*{-0.25in}
\epsscale{1.3}
\plotone{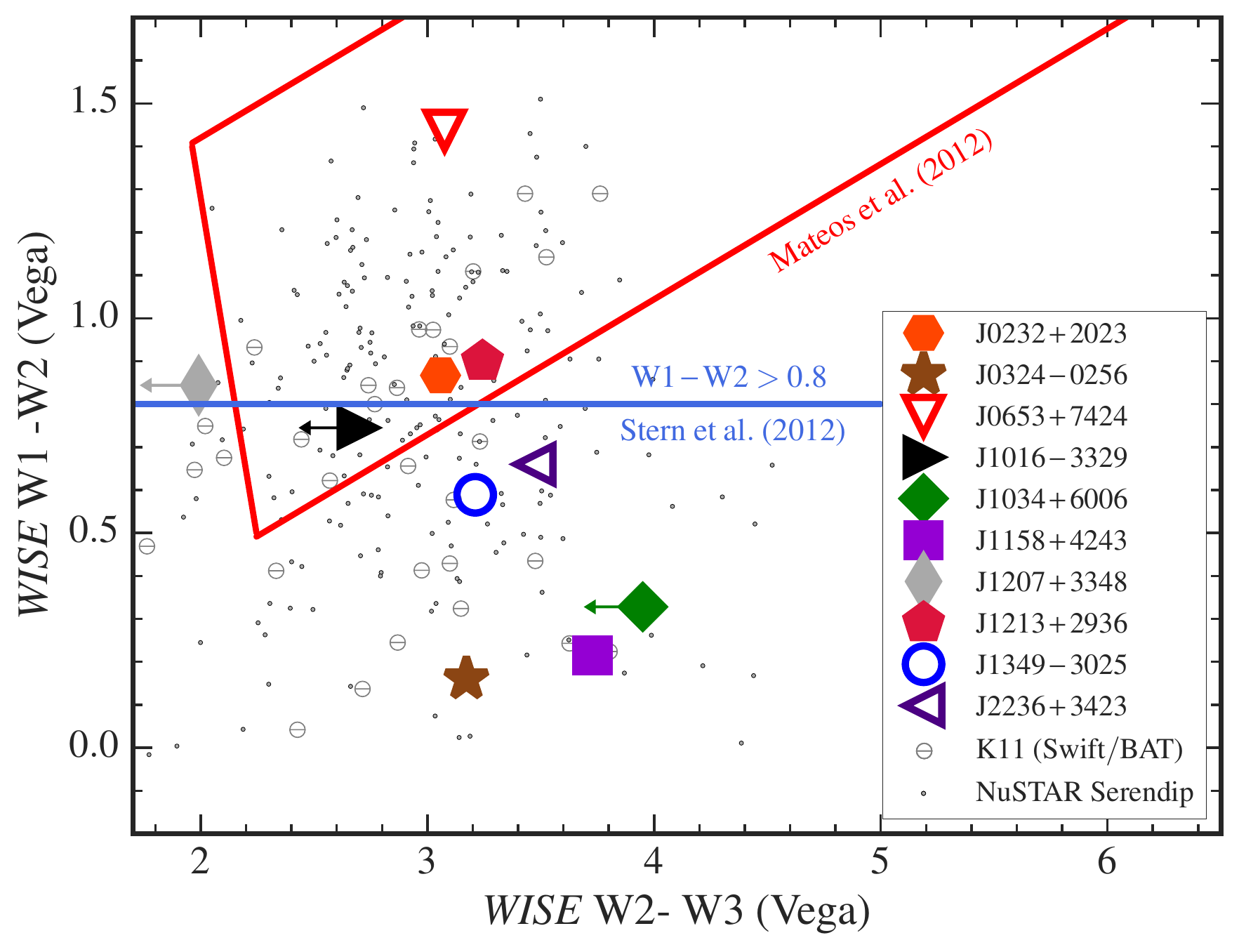}
\caption{{\it WISE} $W1-W2$ versus $W2-W3$ distribution for the \nus\ low-mass AGN sample over-plotted on the \citet[][red wedge]{mate13} and \citet[][blue line]{ster12wise} mid-IR AGN selection criteria. The entire spectroscopic sample of the \nus\ serendipitous survey is shown as gray dots. For comparison, the low-mass galaxies selected from the \swiftbat\ survey \citep{koss11bathost}  are also shown as barred circles (see \S\ref{subsec:xprop} for more details of the \swiftbat\ sample). Large symbols are as in Figure~\ref{fig:rmag_mass}.}
\label{fig:wise}
\end{figure}

We first show the {\it WISE} color-color distribution of our low-mass galaxy sample in Figure~\ref{fig:wise}.
As mentioned earlier, all the \nus\ low-mass galaxies have SNRs higher than 5 in the \textit{W1} and \textit{W2} bands.
There are three objects in our sample with SNR$<5$ in the $W3$ band, and thus we consider their $W2-W3$ colors to be upper limits. 
Only five of the 10 objects in our sample satisfy the \cite{ster12wise} mid-IR AGN selection criteria for luminous AGNs, e.g., $W1-W2 > 0.8$ and $W2<15$. For our sample, the three least X-ray luminous sources all have blue mid-IR colors ($W1-W2 < 0.8$) while the more X-ray luminous AGNs are more likely to lie above the $W1-W2 > 0.8$ criterion. 

By design, the {\it WISE} color-selection criterion has limited selection completeness for low-luminosity AGNs
due to more significant host-galaxy dilution \citep[e.g.,][]{ecka10,donl12}. The fraction of AGNs in our sample with $W1-W2 > 0.8$ is also consistent with the $20-40\%$ mid-IR selection completeness for low-luminosity AGNs in the sample selected using 4.5--10 keV \xmm\ detection \citep{mate13} and the complete \nus\ serendipitous-survey sample (L17).
Due to the low completeness of the $W1-W2 > 0.8$ selection criterion, only one out of the three objects without apparent AGN-powered emission lines has $W1-W2 > 0.8$ (namely J065318+7424.8, see \S\ref{subsec:optspec}). 

Also, two of the most heavily \hbox{X-ray} obscured AGNs (J101609--3329.6 and J115851+4243.2, see Table~\ref{tab:sxprop} and Figure~\ref{fig:wise}) do not meet the $W1-W2>0.8$ color-selection criterion. This suggests that mid-IR color selection methods are not effective in identifying optically normal AGN hosted by low-mass galaxies.
On the other hand, we find that all seven objects in our sample with SNR $>5$ in the \textit{W3} band have red $W2-W3$ colors.
Furthermore, the broad-band SEDs of these objects show that the mid-IR emission at $\gtrsim 6\micron$ is still dominated by the AGN-powered hot-dust component.
With the limited sample size, it is not clear if the red {\it WISE} color at longer wavelengths is
a common feature of hard \hbox{X-ray} selected low-mass AGNs. However, we note that low-mass AGNs selected using other methods, such as the BPT diagram or the presence of the \ion{He}{2} $4686$\AA\ emission line do not show as high a fraction of red $W2-W3$ colors (e.g., Figure 2 and Figure 3 of \citealt{sart15} and Figure 1 of \citealt{hain16}) as our sample does, which might simply be due to the higher redshift and luminosities of our sample. 

\subsection{$M_\bullet-M_\star$ and Eddington ratios}\label{subsec:mbh}
\begin{figure}
\hspace*{-0.25in}
\epsscale{1.3}
\plotone{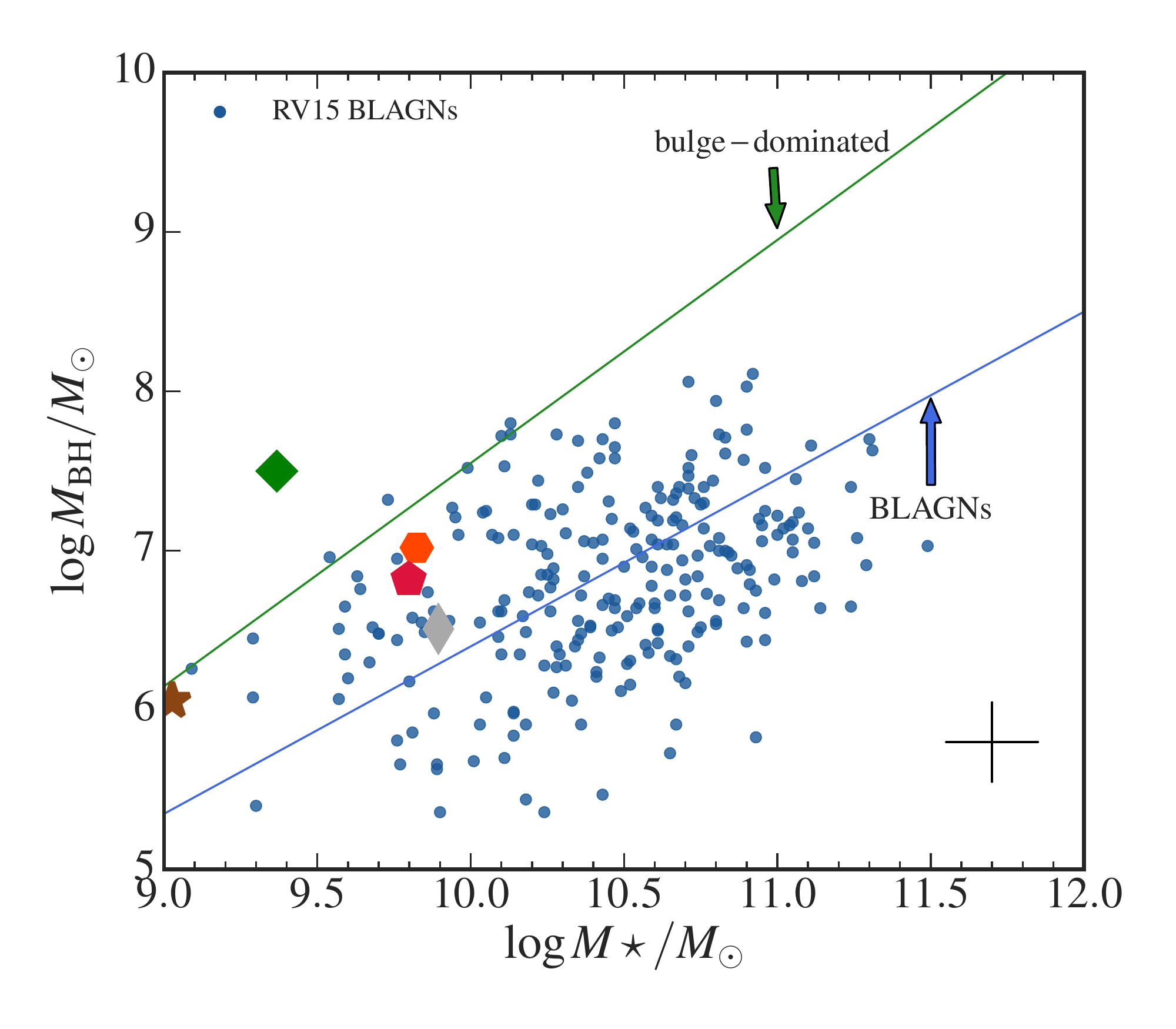}
\caption{$M_\bullet-M_\star$ distribution for the five broad-line AGNs in our sample. For comparison, the $M_\bullet-M_\star$ distribution of the broad-line AGNs presented by \citet[][RV15]{rein15dwarf} are also shown. The best-fit linear relations for the RV15 AGN sample and the bulge-dominated galaxies from \citet[][KH13]{kh13araa} are also shown. Other large symbols are as in Figure~\ref{fig:rmag_mass}. 
The typical $\sim 0.5$ dex systematic uncertainties of virial BH mass estimation \citep[e.g., see][]{rein15dwarf} and the typical $\sim 0.3$ dex uncertainties of the $M_\star$ estimation are shown in the bottom-right corner.}
\label{fig:msig}
\end{figure}

In our sample, five of the seven galaxies shown in Figure~\ref{fig:bpt} exhibit broad $\mathrm{H}\alpha$ emission, allowing estimates of their black-hole masses. The black-hole mass for each broad-line AGN was estimated using Equation 5 of \cite{rein13dwarf} using the $\mathrm{H}\alpha$ line width and luminosity \citep{gh05bhmass}, with the updated BH radius-luminosity relation from \cite{bent13}:
\begin{equation}
\begin{aligned}
&\log (\frac{M_\bullet}{M_\sun}) = \log \epsilon+6.57\\
& + 0.47 \log (\frac{L_{{\rm H}\alpha}}{10^{42} \mathrm{erg\: s}^{-1}}) +
2.06 \log (\frac{{\rm FWHM}_{H_\alpha}}{10^3 \mathrm{km\: s}^{-1}})
\end{aligned}
\end{equation}
Here we choose the constant $\epsilon=1.075$, which is based on the mean virial factor of $\left<f\right>=4.3 $\citep{onke04,grie13a}. 
 We show the $M_\bullet - M_\star$ distribution for the \nus\ low-mass galaxies in Figure~\ref{fig:msig}. 
 For comparison, we also show the best-fit linear relations from \cite{rein15dwarf} for 262 broad-line AGNs in the local universe ($z<0.055$), and for the bulge-dominated galaxies with dynamical $M_\bullet$ measurements from \cite{kh13araa}.\footnote{We adopt the linear $M_\bullet - M_\star$ relation for the bulge-dominated galaxies from \cite{kh13araa} calculated by \cite{rein15dwarf} (i.e., their Equation 6).} 

Overall, the $\log (M_\bullet/M_\sun$) of the five broad-line AGNs in our \textit{NuSTAR}-selected sample ranges from $6.1-7.5$, and thus all are more massive than the \cite{rein15dwarf} relation for broad-line AGNs. 
We also calculate the Eddington ratio for the five broad-line AGNs in our sample using intrinsic $L_{2-10\mathrm{keV}}$ and a constant bolometric correction factor of 22.4 \citep{vasu07bolc}. We find that the Eddington ratio for these AGNs ranges from $3\% - 25\%$. The high values of $M_\bullet$ and Eddington ratio for the five broad-line objects in our sample are likely due to a combination of both the flux limits of \nus\ and the host galaxy dilution effect, as less massive mBHs would either be missed in the current {\it NuSTAR} serendipitous survey or have their broad-line regions buried in the stellar continuum of the host galaxies \citep{hopk09obs}. The Eddington ratios are also listed in Table~\ref{tab:mwprop}.

\subsection{X-ray properties}\label{subsec:xprop}
\begin{figure*}
\epsscale{2.0}
\plottwo{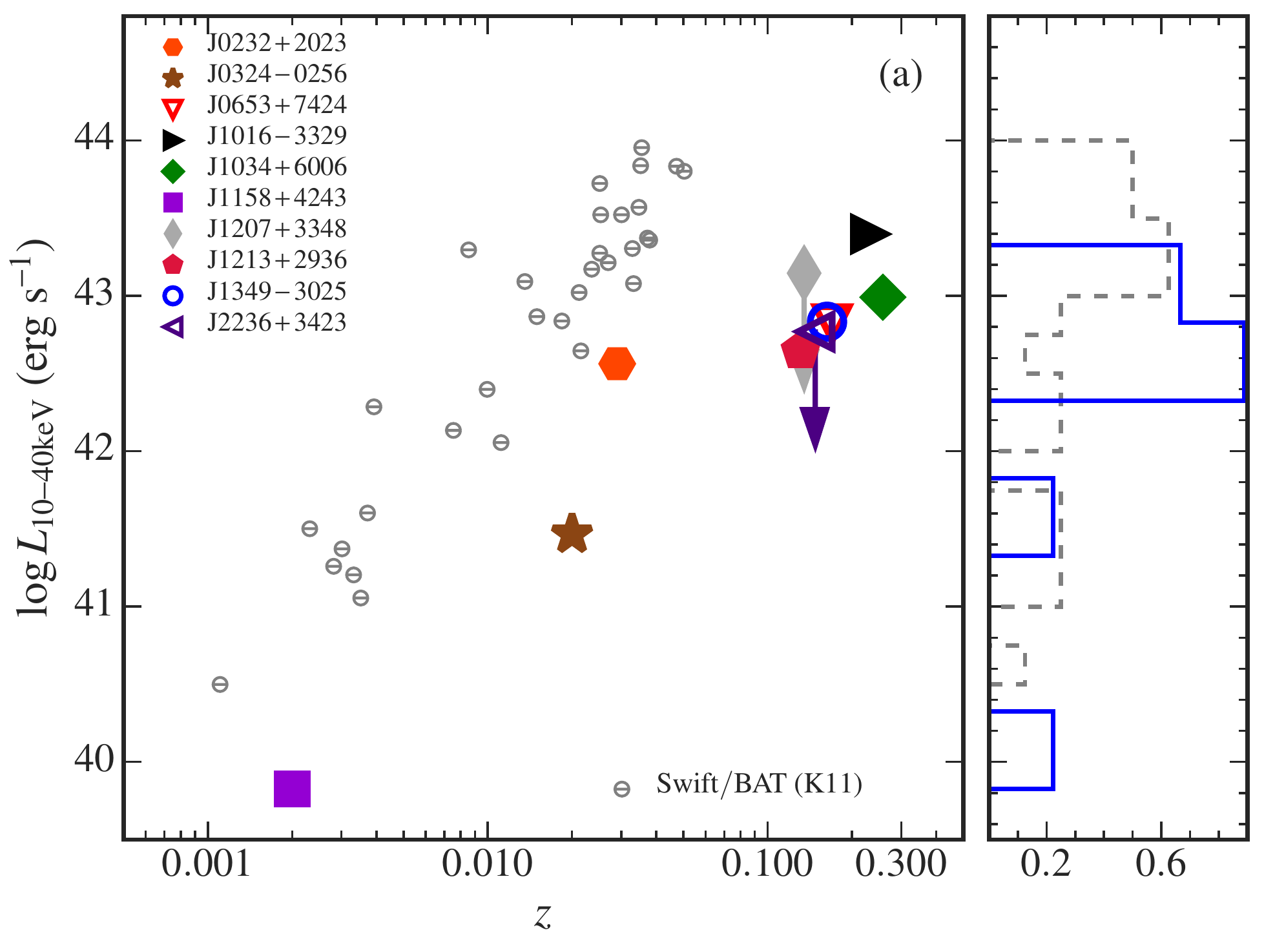}{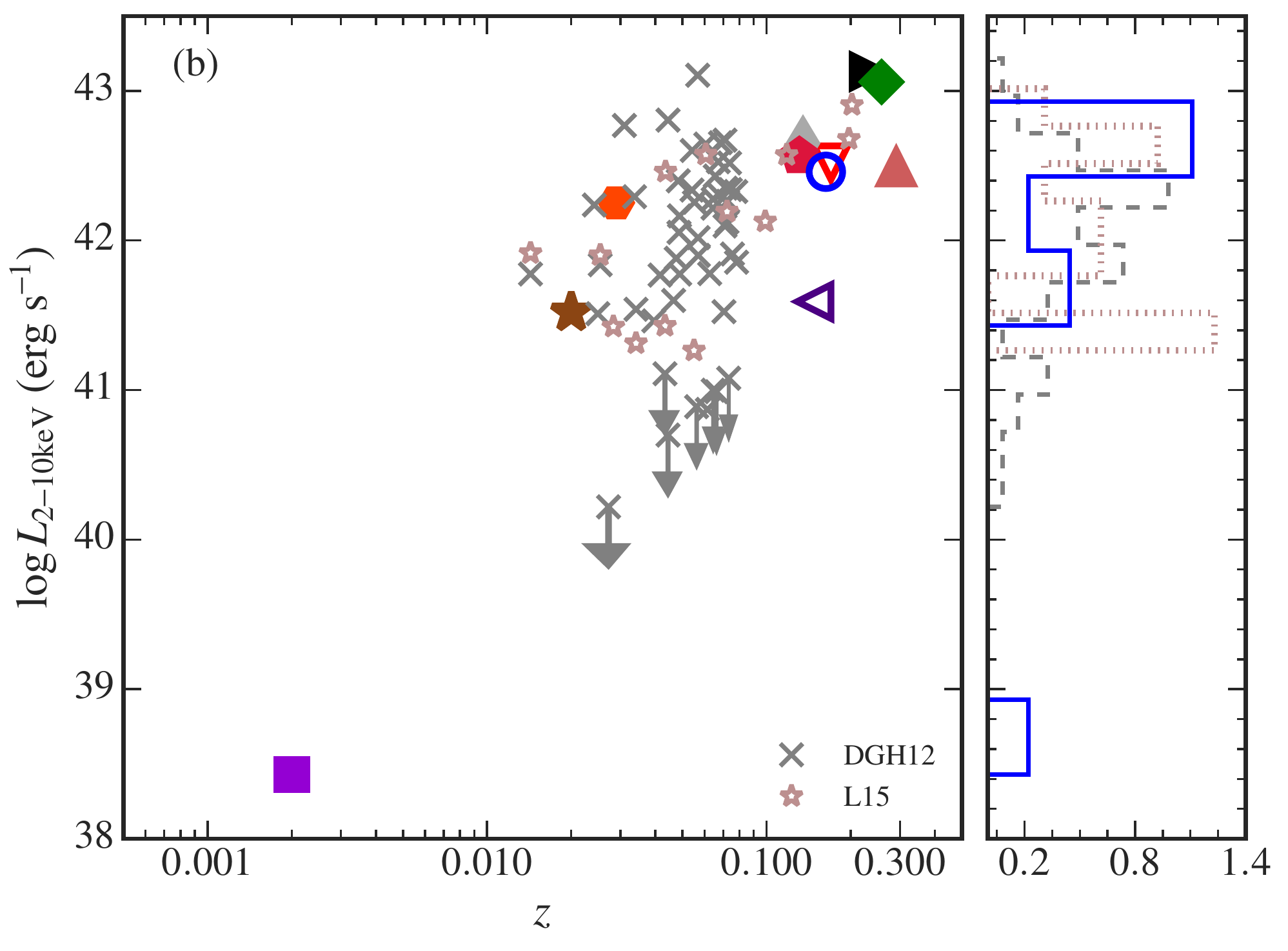}
\caption{(a)-{\it top} : Hard \hbox{X-ray} luminosities of the \nus\ sample and the \swiftbat\ sample from K11 (without correction for absorption).
In the side panel, the normalized histogram of the \nus\ sample is shown as the blue solid line, and the \swiftbat\ sample is shown as the gray dashed line. (b)-{\it bottom} : Rest-frame 2--10 keV \hbox{X-ray} luminosity (not corrected for absorption) distributions of the \nus\ sample
and the optical low-mass AGN samples from \citet[][DGH12]{dong12xray} and \citet[][L15]{ludl15}.
In the side panel, the normalized histogram of the \nus\ sample is shown as the solid blue line,
and the DGH12 and L15 samples are shown as the dashed and dotted lines, respectively. }
\label{fig:lx_z}
\end{figure*}

Here we compare the \hbox{X-ray} properties of our sample with those of existing low-mass AGNs reported in the literature.
We emphasize that due to the challenges of detecting faint \hbox{X-ray} emission from mBHs,
existing low-mass AGNs with \hbox{X-ray} observations are limited to a small number of low-redshift ($z\lesssim 0.2$) objects,
and an even smaller number of higher redshift objects \citep[$z\lesssim 1$, e.g.][]{schr13,pard16} found in deep-survey regions.
Since our sample is limited to $z\lesssim 0.3$, we assess our results by comparing the properties of our sample
with two samples of low-mass AGNs in the local universe: the low-mass AGNs detected by {\it Swift}/BAT,
and those selected using broad optical emission lines.

We select the low-mass AGNs from the \swiftbat\ AGNs studied in \citet[][K11 hereafter]{koss11bathost},
which is a sample of 185 AGNs selected from the 22-month and 58-month {\it Swift}/BAT all-sky surveys.
The sensitivity limit of the K11 sample is $(1.1-1.5)\times10^{-11}$ erg cm$^{-2}$ s$^{-1}$ in the $14-195$ keV band,
which is equivalent to $(0.6-0.8)\times10^{-11}$ erg cm$^{-2}$ s$^{-1}$ in {\it NuSTAR}-FB,
assuming a typical AGN photon index of $\Gamma=1.8$.
K11 also estimated the stellar mass for each AGN in their sample,
which makes it possible for us to select AGNs with $M_\star$ similar to our low-mass \nus\ sample.
There are a total of 32 galaxies from K11 that could be considered as ``low-mass'' ($M_\star\lesssim 10^{10}M_\sun$) similar to our \nus\ sample. 
We note that the approach used to estimate the host-galaxy photometry and stellar mass is different in K11 compared to our approach outlined in \S\ref{sec:data}.
However, for simplicity, we directly adopt the values provided by K11
since we find that the $M_\star$ values derived using their approach and ours have little systematic difference (see Appendix B for a comparison between the quantities calculated using our approach and those directly obtained from K11). \par

For AGNs selected from low-mass galaxies based on the presence of AGN-powered optical emission lines, there have been a number of studies focusing on objects selected from the SDSS \citep[e.g.,][]{gh07imbh,bart08,dong12imbh,rein13dwarf,mora14dwarf,sart15}, but only a fraction of these optical AGNs has soft \hbox{X-ray} follow-up observations.
We focus on the two largest optical low-mass AGN samples with \hbox{X-ray} follow-up observations:
the 50 AGNs with \chandra\ 2 ks snapshot observations discussed in \cite{dong12xray}
and the 14 AGNs with deeper \xmm\ observations ($> 10$ ks) from \cite{ludl15}. Both samples are selected from the 229 low-mass AGNs identified using SDSS DR4 \citep{gh07imbh}.

In Figure~\ref{fig:lx_z}a, we compare the $L_{10-40\mathrm{keV}}$ values of our \nus\ sample with those of the 32 low-mass galaxies selected from K11.
The median $\log (L_{10-40\mathrm{keV}}$/erg s$^{-1}$) is $42.45\pm0.31$ for our \nus\ sample and $42.88\pm0.15$ for the K11 subsample.
The uncertainties for these median values were estimated using a bootstrapping analysis in which we randomly draw the samples with replacement 100 times.
We also show the $L_{2-10\mathrm{keV}}$ distributions of our \nus\ sample and the optical low-mass AGN samples in Figure~\ref{fig:lx_z}b.
The median $\log (L_{2-10\mathrm{keV}}$/erg s$^{-1}$) is $42.48 \pm 0.19$ for the \nus\ sample and $42.10\pm0.09$ for the two optical samples.

We note that the redshift of the K11 sample is limited to  $z<0.05$. However, beyond $z=0.05$,
it is not likely for AGNs powered by accretion onto mBHs to emit hard \hbox{X-ray} emission exceeding the \swiftbat\ sensitivity limit of $\approx10^{-11}$ \hbox{ergs~cm$^{-2}$~s$^{-1}$.}
For the {\it NuSTAR}-selected sample in this work, the average redshift is 0.14, and the $L_{10-40\mathrm{keV}}$ of the \nus\ sample is $\approx 0.4$ dex fainter than for the low-mass galaxies found in the K11 sample. 
This again highlights the excellent sensitivity of \nus\ and its ability for studying low-mass galaxies beyond the local universe.

For the optically selected AGNs, \cite{dong12xray} also targeted low-mass AGNs at low redshift ($z\lesssim 0.08$)
due to the flux limit of their 2 ks \chandra\ snapshot observations.
For the \cite{ludl15} sample, the deeper \xmm\ observations reach $z\lesssim 0.2$,
which is more comparable to our \nus\ sample. In Figure~\ref{fig:lx_z}b,
we find that our \nus\ sample occupies a similar region of $L_{2-10\mathrm{keV}}-z$ parameter space to the \hbox{X-ray} follow-up observations of optical AGNs with broad emission lines and $M_\bullet \approx 10^6 M_\sun$,
demonstrating the strength of serendipitous \nus\ observations 
in detecting hard \hbox{X-ray} emission from 
low-mass AGNs in the low-redshift universe.

\begin{figure}
\epsscale{1.2}
\plotone{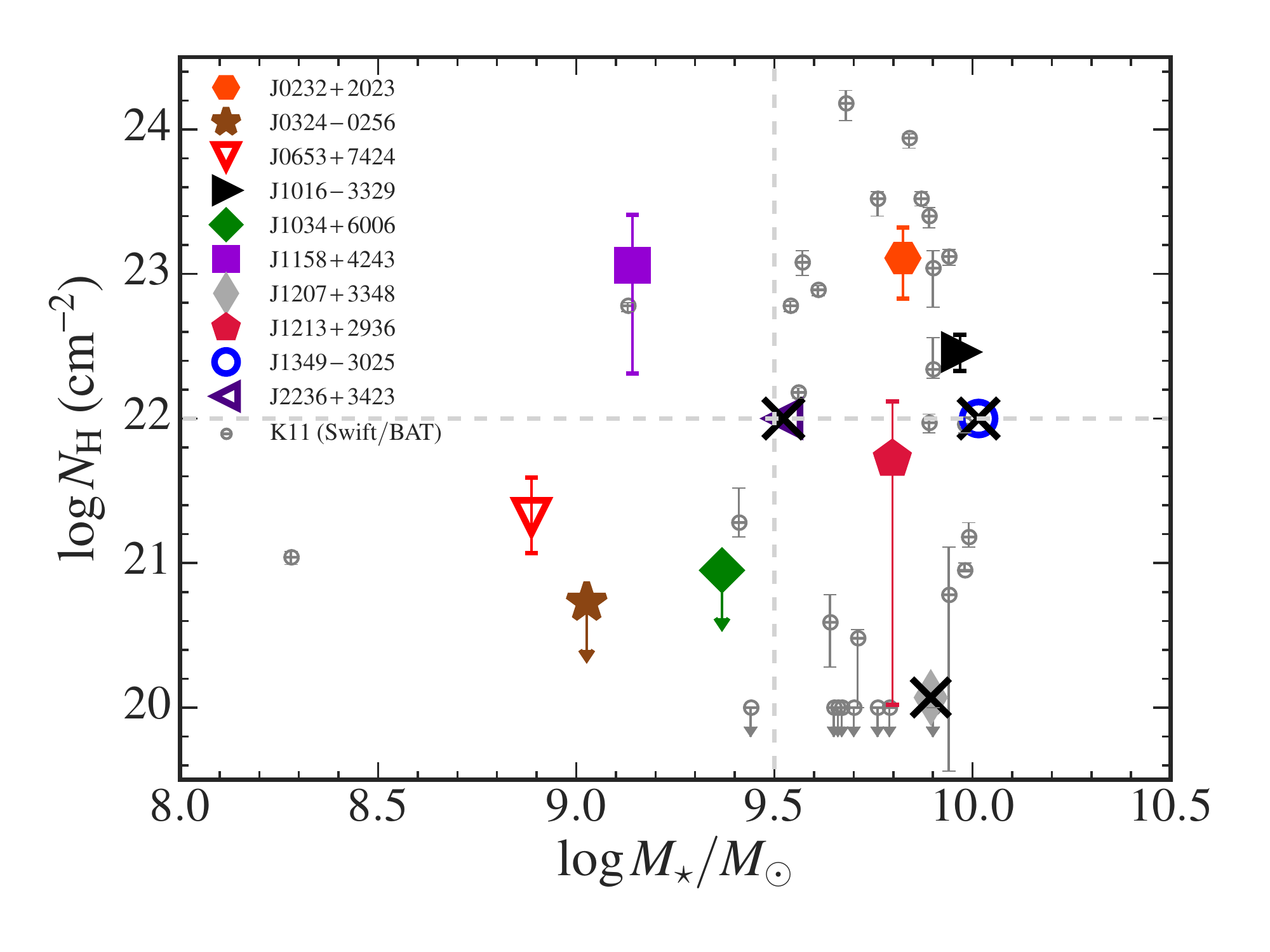}
\caption{$N_{\rm H}$ versus $M_\star$ for low-mass AGNs selected using \nus\ and \textit{Swift}/BAT (gray dots).
The $N_{\rm H}$ values derived based on X-ray spectral fitting are shown as the symbols with vertical error bars that correspond to $90\%$ confidence interval. Sources with $N_{\rm H}$ values estimated based on their hardness ratio are shown as crossed symbols without vertical error bars. See \S\ref{subsec:softxray} for details. The typical uncertainty in $M_\star$ is 0.3 dex (see \S\ref{subsec:sedfitting}).
We note that there are only two $N_{\rm H} > 10^{22}$ cm$^{-2}$ AGNs hosted by dwarf galaxies with $\log M_\star/M_\sun < 9.5$,
which highlights the limited sample size of \hbox{X-ray} obscured AGNs in dwarf galaxies.}
\label{fig:nh_mstar}
\end{figure}

Some of the \nus\ low-mass AGNs reported in this work are \hbox{X-ray} obscured (see \S\ref{subsec:softxray}). We note that there is not yet a clear understanding of the obscured AGN population hosted by low-mass galaxies,
which is primarily due to the existing \hbox{X-ray} observations mostly having targeted AGNs with broad emission lines,
and there are few \hbox{X-ray} selected AGNs hosted by low-mass galaxies that are obscured to the best of our knowledge.
With the {\it NuSTAR}-selected sample and the low-mass AGNs from K11, we can take a first step in constraining the \hbox{X-ray} obscured fraction using hard \hbox{X-ray} selected low-mass AGNs, 
although caution is required because the \nus\ serendipitous survey is relatively shallow (with a median exposure time of 28 ks, see L17) and might still not be able to detect efficiently low-mass Compton-thick AGNs.
We utilize the \hbox{X-ray} spectral-analysis results for the \nus\ sources and the ancillary soft \hbox{X-ray} data
for the K11 AGNs from the literature (see Appendix~\ref{appendix:bat_k11} and Table~\ref{tab:bat}).
For \hbox{X-ray} detected AGNs, $N_{\rm H} = 10^{22}$ cm$^{-2}$ is a commonly used value for separating \hbox{X-ray} type 1 and type 2 objects
\citep[but see also][for the use of slightly lower $N_{\rm H}$ values for classifying type 2 objects]{merl13}.
We adopt this criterion to select X-ray type 2 AGNs from the \nus\ and \swiftbat\ samples. We find that the \hbox{X-ray} obscured fraction for the seven objects in the \nus\ sample with $N_{\rm H}$ measurements is $43^{+15}_{-18}\%.$\footnote{We use the \cite{came13} method to calculate the $68.3\%$ binomial confidence limits of the obscured fraction.} This obscuration is likely a lower-limit as some of the most heavily obscured AGNs would not have been detected even with {\it NuSTAR} \citep[e.g.][]{ster14nustar,lans14nustar}. For the K11 low-mass AGNs, the obscured fraction is $51\pm 8\%$. For the combined sample of 42 AGNs, we compute the obscured fraction to be $47^{+8}_{-7}\%$, but we caution that the two samples have different selection functions. The $N_{\rm H}$ vs. $M_\star$ distributions for our sample and the K11 low-mass AGNs are shown in Figure~\ref{fig:nh_mstar}.

While the accuracy of the obscured fraction presented here is limited by the small sample sizes and selection functions
of the \nus\ and K11 low-mass AGN samples, the results from the two samples are both broadly consistent with the findings of
previous studies of the luminosity-dependent AGN obscured fraction. For example, the obscured fraction derived using Equation 5 from \citet{luss13} is $\sim 50\%$ for the median luminosity of our AGN sample. 
When we focus only on the nine dwarf galaxies with $M_\star<3\times10^9M_\sun$,
there are only two \hbox{X-ray} obscured AGNs (i.e., $N_{\rm H}>10^{22}$ cm$^{-2}$) from the combined \nus\ and K11 sample, which is equivalent to $22.2^{+18.3}_{-8.1}\%$.
To investigate this possible reduction of the obscured fraction, we separate the combined hard \hbox{X-ray} AGN sample into two subsamples,
the ``low-mass galaxies'' with $3\times10^{9} M_\odot < M_\star < 10^{10}M_\odot$,
and the ``dwarf galaxies'' with $M_\star<3\times 10^{9}M_\odot$.
We apply a Peto-Prentice test on the $N_{\rm H}$ distributions for both subsamples 
to account for the upper-limits on the $N_{\rm H}$ values \citep[e.g., see Astronomy SURVival Analysis,][]{soft_asurv}. 
The two-sample Peto-Prentice probability for both samples to follow the same distribution is $17.0\%$.

\section{Discussion and conclusions}
In this work, we present the first {\it NuSTAR}-selected sample of 10 low-mass galaxies harboring hard X-ray-emitting AGNs
from the 40-month \nus\ serendipitous survey (L17).
Compared to low-mass AGNs in the previous-generation hard \hbox{X-ray} \swiftbat\ survey,
our {\it NuSTAR}-selected sample includes several low-redshift objects with much lower hard \hbox{X-ray} luminosities as well as low-mass AGNs at higher redshifts. The soft \hbox{X-ray} luminosities of our objects are
consistent with those of optically selected low-mass AGNs in the low-redshift universe.
We find that $30^{+17}_{-10}\%$ of the \nus\ AGNs in our sample do not have AGN-powered high-ionization lines in their optical spectra,
which demonstrates the capability of \nus\ for detecting a low-mass AGN population that might be missed in wide-area optical surveys.

We also study the {\it WISE} colors of our sample and find that the majority of the \hbox{X-ray} obscured AGNs
and optically normal AGNs in our sample do not have red $W1-W2$ colors similar to those of luminous mid-IR AGNs.
We note that eight of the ten low-mass \nus\ galaxies still show red mid-IR colors at longer mid-IR wavelengths ($W2-W3>3$) that are likely due to the presence of hot dust powered by AGNs. However, a substantial number of low-mass star-forming galaxies also exhibit red {\it WISE W2$-$W3} color \citep{hain16}, suggesting that the effectiveness of using only the red {\it WISE} colors to select low-mass AGNs missed by optical surveys is limited.

We constrain the obscured fraction of hard \hbox{X-ray} selected low-mass AGNs by combining our sample with the K11
{\it Swift}/BAT low-mass AGNs. We find that hard \hbox{X-ray} selected low-mass AGNs have an ``obscured fraction'' of $47^{+8}_{-7}\%$. This is consistent with the obscured fraction extrapolated from studies of the luminosity dependent fraction of obscured AGNs \citep[e.g., Equation 5 of][]{luss13}. However, when focusing on the ``dwarf galaxies'' ($M_\star<3\times 10^9M_\sun$), the fraction of AGNs with $N_{\rm H}>10^{22}$~cm$^{-2}$ drops to $22.2^{+18.3}_{-8.1}\%$ (or 2/9).

Notably, the two heavily obscured AGNs hosted by dwarf galaxies (J115851+4243.2 from our sample and J0505.8--2351 from K11) do not have blue host-galaxy colors similar to the high-redshift galaxies with stacked \hbox{X-ray} spectra that suggest the presence of heavily obscured AGNs \citep{xue12cxb,mezc15}. 
The likely reason is that the obscured AGN population suggested by the high-redshift \hbox{X-ray} stacking studies of star-forming galaxies is still less luminous (e.g., $\langle L_{2-10\mathrm{keV}}\rangle \approx 4.8\times10^{40}$ erg s$^{-1}$ for ``sample D'' in \citealt{xue12cxb}, and $L_{2-10\mathrm{keV}}\lesssim 6\times 10^{40}$ erg s$^{-1}$ for the \citealt{mezc15} sample) than the majority of our sample and thus are not detected in the \nus\ serendipitous survey. 

On the other hand, several recent studies have found that AGNs in dwarf galaxies selected based on optical emission-line ratios
have a high Seyfert 2 fraction \citep[e.g.,][]{rein13dwarf,mora14dwarf}. Some studies have also suggested that low-luminosity Seyfert 2 galaxies are unobscured and lack broad line regions due to falling below a critical accretion luminosity that is independent of the Eddington rate of the accreting mBH
\citep[e.g.,][]{bian08,trum11,elit14,elit16}
Therefore, \hbox{X-ray} observations are still essential to determine whether a low-luminosity AGN is obscured by intervening gas/dust. For AGNs that are more heavily obscured, hard X-ray observations provide arguably the best constraint on whether a low-luminosity AGN is obscured. 
Although the current sample size of hard \hbox{X-ray} selected AGNs hosted by dwarf galaxies is still limited, 
our results have demonstrated the capability of \nus\ in detecting heavily obscured AGNs hosted by dwarf galaxies. 
We note that the current serendipitous-survey catalog (L17) from which our low-mass AGN sample is drawn is based on the 40-month \nus\ observations with $\approx 50\%$ optical spectroscopic coverage. 
Therefore, a 10-year \nus\ serendipitous survey with complete optical spectroscopic follow-up observations will likely increase the sample size of {\it NuSTAR}-detected AGNs in low-mass galaxies by more than a factor of five compared to what is presented here. 
But even a 10-year serendipitous survey may have only a few heavily obscured AGNs (similar to our J115851+4243.2), motivating targeted follow-up {\it NuSTAR} observations of heavily obscured AGNs selected at other wavelengths to build a more complete picture of the AGN population in dwarf galaxies.

In conclusion, this small sample of {\it NuSTAR}-selected low-mass AGNs has demonstrated that \textit{NuSTAR} is capable
of detecting a variety of AGNs in low-mass galaxies that are complementary to the existing emission-line low-mass AGNs
that are found in optical surveys and previous-generation hard \hbox{X-ray} surveys.
We stress that the spectroscopic observations and ancillary soft \hbox{X-ray} data are instrumental in the construction
of our {\it NuSTAR}-selected sample. With the small volume of the current \nus\ surveys,
the most-efficient method of systematically searching for low-mass AGNs that are not broad emission-line AGNs
may still be cross-matching soft \hbox{X-ray} observations with optical spectroscopic surveys of galaxies.
However, we note that there are only three objects in our sample with SDSS spectra, which is due to the flux limits (${\it r}<17.77$
for the main galaxy targets) for spectroscopic observations by the SDSS.
Moreover, a recent study that matched the \chandra\ Source Catalog with local dwarf galaxies in the SDSS is limited to the local universe and has primarily found low-luminosity \hbox{X-ray} sources that are not likely to be only associated with AGN activity \citep{lemo15}.
With a greatly improved survey volume, the upcoming eROSITA all-sky \hbox{X-ray} survey \citep{inst_erosita}
and next-generation wide-area spectroscopic surveys such as the {\it Subaru} PFS survey \citep{survey_pfs} should reveal
many more \hbox{X-ray} AGNs hosted by dwarf galaxies at moderate redshifts, as these surveys will reach flux limits that are deep enough to recover the majority of sources similar to the \nus\ objects reported in this work. However, as some of the targets in our sample are heavily obscured, \nus\ remains a key observatory for providing insights about low-mass AGNs
that are obscured by Compton-thick column densities similar to the megamaser AGNs.

\acknowledgments{
We thank the referee for carefully reading the manuscript and providing helpful comments.
This work was supported under NASA contract No. NNG08FD60C, and made use of data from the {\it NuSTAR} mission, a project led by the California Institute of Technology, managed by the Jet Propulsion Laboratory, and funded by the National Aeronautics and Space Administration. We thank the {\it NuSTAR} Operations, Software and Calibration teams for support with the execution and analysis of these observations. This research has made use of the {\it NuSTAR} Data Analysis Software (NuSTARDAS) jointly developed by the ASI Science Data Center (ASDC, Italy) and the California Institute of Technology (USA).
C-T.J.C. and W.N.B. acknowledge support from Caltech {\it NuSTAR} subcontract 44A-1092750. Support for A.E.R. was provided by NASA through Hubble Fellowship grant HST-HF2-51347.001-A awarded by the Space Telescope Science Institute, which is operated by the Association
of Universities for Research in Astronomy, Inc., for NASA, under contract NAS 5-26555. 
D.M.A gratefully acknowledges support from Science and Technology Facilities Council (ST/L00075X/1).
F.E.B. and C.R. acknowledge support from NASA NuSTAR A01 Award NNX15AV27G, CONICYT-Chile grants Basal-CATA PFB-06/2007, FONDECYT Regular 1141218 and 1151408, China-CONICYT Fellowship, and the Ministry of Economy, Development, and Tourism's Millennium Science Initiative through grant IC120009, awarded to The Millennium Institute of Astrophysics, MAS.
This publication makes use of data products from the Two Micron All Sky Survey,
which is a joint project of the University of Massachusetts and the Infrared Processing and Analysis Center/California Institute of Technology,
funded by the National Aeronautics and Space Administration and the National Science Foundation.
Funding for SDSS-III has been provided by the Alfred P. Sloan Foundation, the Participating Institutions,
the National Science Foundation, and the U.S. Department of Energy Office of Science. The SDSS-III website is \url{http://www.sdss3.org/}.} This research has made use of ``Aladin sky atlas'' developed at CDS, Strasbourg Observatory, France \citep{soft_aladin}. This work has also made use of observations made with the Spitzer Space Telescope, obtained from the NASA/ IPAC Infrared Science Archive, both of which are operated by the Jet Propulsion Laboratory, California Institute of Technology under a contract with the National Aeronautics and Space Administration.
\facility{NuSTAR, Chandra, XMM-Newton, Swift/BAT, Swift/XRT, WISE, Keck, NTT, Palomar}

\software{Astropy, CIAO, HEAsoft, XMMSAS, XSPEC}

\appendix

\section{Notes on individual galaxies}
Here we briefly summarize the multiwavelength properties and describe the procedures of \hbox{X-ray} spectral analysis for each object in our sample. The resulting \hbox{X-ray} and optical spectra, and other relevant information for each object are shown in the online figure set (Figure~\ref{fig:figset}). 

\begin{figure}
\epsscale{1.2}
\plotone{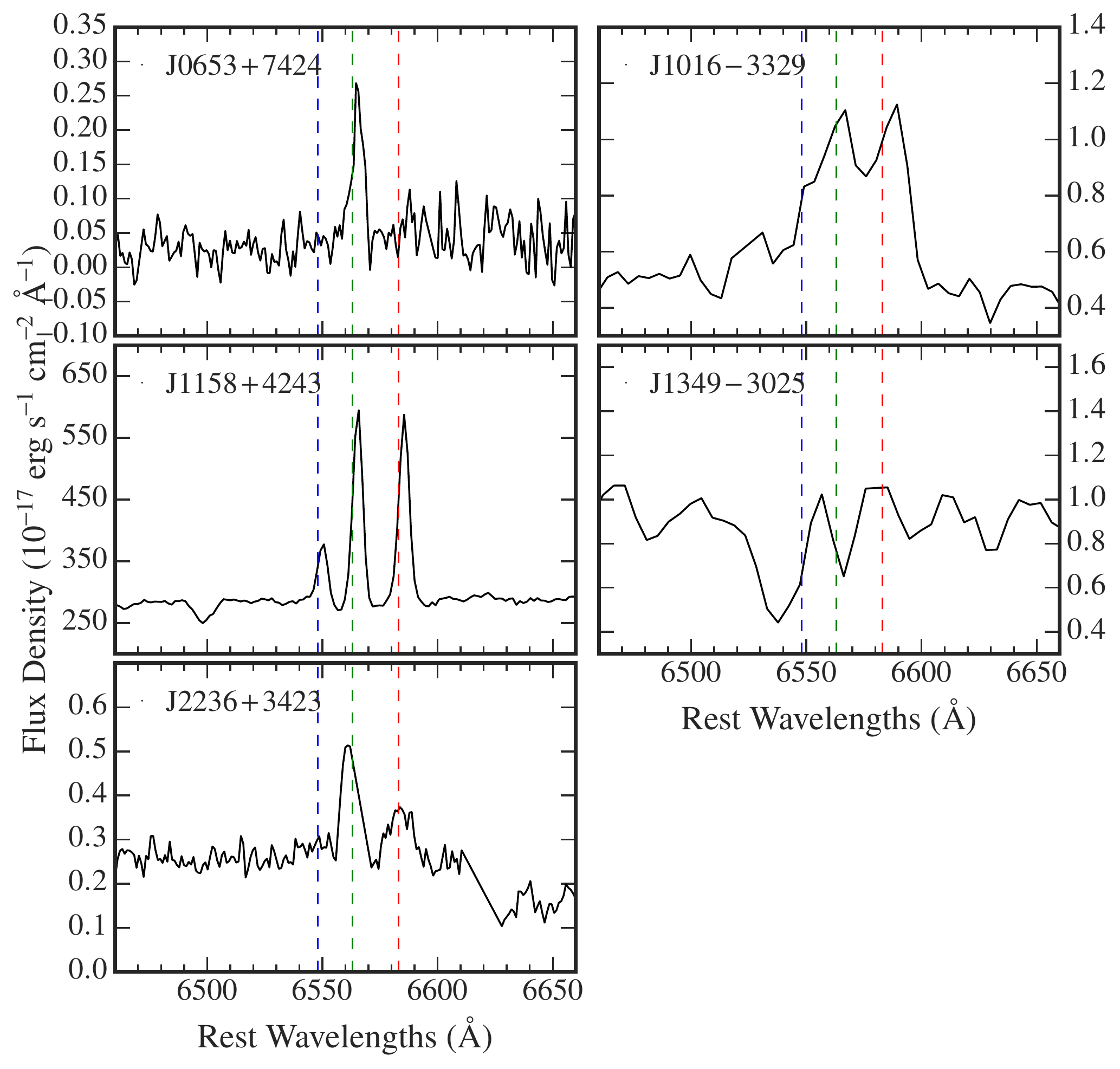}
\caption{The spectra for objects without broad H$\alpha$ lines plotted in the same rest-frame wavelength range as Figure~\ref{fig:halpha}. 
In each panel, the emission lines [\ion{N}{2}]6548\r{A}, H$\alpha$ 6563\r{A}, and [\ion{N}{2}] 6583\r{A} are marked as the blue, green, and red dashed lines, respectively. We note that J101609--3329.6 and J115851+4243.2 still have strong narrow H$\alpha$ lines, but the best-fit broad H$\alpha$ component for these two objects does not exceed the 500 km s$^{-1}$ threshold we used to identify broad-line AGNs. For J065318+7424.8 and J134934-3025.5, the rest-frame wavelength for H$\alpha$ emission lines coincides with the telluric A-band lines. }
\label{fig:noha}
\end{figure}

\paragraph{J023229+2023.7}
This galaxy has the optical spectrum of a typical broad-line AGN. However, the \nus\ spectrum suggests the presence of a substantial amount of absorption in the X-rays. 
For both available \xmm\ observations, the position of J023229+2023.7 is located near the \xmm\ chip gap of the PN detector and on the dead CCD of the MOS1 detectors, and they are both severely affected by background flaring. Data from the MOS2 detector was taken under the timing mode and there is no useful spectral information. 
We only consider the \nus\ data for \hbox{X-ray} spectral analysis. We first fit the data with an unabsorbed power-law. The resulting power-law photon index of $\Gamma=1.2$ and the residuals at $\approx 6.4$ keV suggest the presence of absorption and an Fe K$\alpha$ line component. Thus, we fit the spectrum again with an absorbed power-law model with an Fe K$\alpha$ line at rest-frame 6.4 keV. The intrinsic photon index is fixed at 1.8 to stop convergence at low values due to the degeneracy with $N_{\rm H}$. The resulting column density is $1.4^{+0.8}_{-0.6}\times 10^{23}$ cm$^{-2}$, and the Fe K$\alpha$ line equivalent width is $\approx0.5$ keV.
We note that the significant equivalent width of the Fe K$\alpha$ line and the residuals above $>20$ keV may imply that J023229+2023.7 has an additional Compton-reflection component. \par

\paragraph{J032459--0256.2} This object has already been identified as a dwarf galaxy in \cite{alex13nustar}
with $M_\star=2\times10^{9}M_\sun$. Its soft \hbox{X-ray} luminosity, $L_{2-10\mathrm{keV}} = 3.3\times 10^{41}$ erg s$^{-1}$,
is derived by jointly fitting the available \nus\ and \swiftxrt\ data.
We find that the \hbox{X-ray} spectrum of J032459--0256.2 is consistent with an absorbed power-law with $\Gamma = 1.9^{+0.3}_{-0.2}$
and intrinsic $N_{\rm H} < 3.4\times10^{21}$ cm$^{-2}$. Although its 2--10 keV \hbox{X-ray} luminosity does not exceed the $10^{42}$ \hbox{erg s$^{-1}$} threshold, its optical spectrum exhibits strong broad H$\alpha$ and narrow high-ionization emission lines powered by an AGN. The BH mass of this object is $\log M_\bullet/M_\sun = 6.06$ based on its broad
H$_\alpha$ emission. With the clear optical AGN signatures, we consider J032459--0256.2 to be a {\it bona fide} dwarf AGN.

\paragraph{J065318+7424.8} \nus\ J065318+7424.8 is a faint emission-line galaxy with $\log M_\star/M_\sun= 8.9$. 
Since this object has sufficient photons from both \nus\ and \chandra, we jointly fit the \nus\ and \chandra\ data.
We find that its \hbox{X-ray} spectrum is consistent with an absorbed power-law with modest intrinsic column density
($N_{\rm H}=2.1^{+1.8}_{-1.2}\times10^{21}$ cm$^{-2}$), 
suggesting the lack of AGN-powered emission lines is not due to the presence of heavy obscuration. The soft \hbox{X-ray} luminosity for this object is $L_{2-10\mathrm{keV}}\approx 3.0\times 10^{42}$ erg s$^{-1}$, suggesting that J065318+7424.8 is indeed powered by accretion onto a massive BH. 

\paragraph{J101609--3329.6} %ic2560_S2
Based on its optical spectrum, J101609--3329.6 is a narrow-line Seyfert galaxy (type 2). 
To explore if its \hbox{X-ray} properties are consistent with those of a type 2 AGN, we first examine the archival \chandra\ data of J101609--3329.6. The $10$ ks \chandra\ image reveals that there are two point sources with $\approx 5\arcsec$ separation, and only one of the point sources has significant $>2$ keV \hbox{X-ray} photons. Thus, we consider the \chandra\ point source with more hard \hbox{X-ray} photons to be the correct counterpart of the AGN in J101609--3329.6. Although the \chandra\ photon counts are too low for spectral analysis, the \chandra\ 2--8 keV to 0.5--2 keV photon-counts ratio suggests that the AGN in J101609--3329.6 is obscured. We next extracted the \xmm\ pn spectrum for J101609--3329.6 with a $10\arcsec$ radius and jointly fit the \nus\ data with the \xmm\ data. The $2-24$ keV spectrum could be fitted with an absorbed power-law with $\Gamma=1.28\pm0.3$ and $N_{\rm H}\approx 2.9\pm1.0\times10^{22}$ cm$^{-2}$, supporting the indication from the \chandra\ band ratio and the lack of broad emission lines in the optical spectrum. 

\paragraph{J103410+6006.7}
J103410+6006.7 is also an optical broad-line AGN. The \nus\ photon counts of J103410+6006.7 are not sufficient for spectral fitting. Therefore, we only consider the \xmm\ data for this object. The \xmm\ spectrum is consistent with a typical unabsorbed AGN with $\Gamma=1.7\pm0.2$ and $N_{\rm H} < 8.9\times10^{20}$ cm$^{-1}$ (90\% upper limit). Among the broad-line AGNs in our \nus\ low-mass galaxy sample, J103410+6006.7 has the highest black hole to galaxy mass ratio, which places J103410+6006.7 in the 
$M_\bullet-M_\star$ region occupied by the local bulge-dominated galaxies.

\paragraph{J115851+4243.2} J115851+4243.2, also known as IC~750, is the lowest redshift galaxy in our sample ($z=0.002$).
The $L_{2-10\mathrm{keV}}=2.7\times 10^{38}$ erg s$^{-1}$ is calculated from the publicly available on-axis \chandra\ observation.
In detail, we first visually inspected the publicly available \chandra\ and \xmm\ images and found that J115851+4243.2 appears to be extended in both \chandra\ ($0.5-8$ keV) and \xmm\ ($0.2-12$ keV) images. Considering the higher-resolution \chandra\ image, we find that the $>2$ keV \hbox{X-ray} emission primarily originates from a central $\approx2\arcsec$ region that coincides with the SDSS fiber location ($< 0.3\arcsec$ separation).
There are also two additional off-nuclear \hbox{X-ray} sources $\approx 5\arcsec$ away from the optical position.
In the lower energy bands ($E<2$ keV), the \chandra\ image appears more extended. Considering the \xmm\ image, the central nucleus cannot be distinguished from the off-nuclear point sources seen by {\it Chandra}. This is likely the reason that the \xmm\ $3-8$ keV flux is higher than the \chandra\ $3-8$ keV flux (see Table~\ref{tab:sxprop}) by a factor of $\sim 1.8$. 
Thus, we extract the \chandra\ spectrum from the central $2\arcsec$ region to avoid possible contamination from off-nuclear \hbox{X-ray} binaries. 
The \chandra\ spectrum still requires an additional diffuse thermal plasma component (we use the {\tt XSPEC} VMEKAL model in addition to an absorbed power-law, i.e. the following XSPEC model: {\sc tbabs$\times$(tbabs$\times$vmekal+tbabs$\times$zpow)}) to have an acceptable fit of the 0.5--8 keV spectrum. With a fixed photon index $\Gamma=1.8$, we obtain a best-fit column density of $1.2^{+1.4}_{-1.0}\times10^{23}$ cm$^{-2}$ ($\chi^2=58.27/71$), 
suggesting the AGN in J115851+4243.2 is heavily obscured and has an intrinsic X-ray luminosity of $L_{2-10\mathrm{keV}}^{\mathrm{int}} = 5.4\times10^{40}$ erg s$^{-1}$. \par

With the low $L_{2-10\mathrm{keV}}$ of J115851+4243.2, the best-fit line-of-sight column density is still insufficient to promote the intrinsic \hbox{X-ray} luminosity to the range of typical AGNs. However, as we have briefly discussed in \S\ref{subsec:optspec}, the optical emission line ratios of J115851+4243.2 strongly suggest the presence of an AGN. Even though the \hbox{X-ray} luminosity of J115851+4243.2 is low, it is not likely for a typical \hbox{X-ray} binary to produce ionizing photons that would push the emission-line ratios of J115851+4243.2 to the observed values (see Figure~\ref{fig:bpt}). Moreover, the angular proximity of the \chandra\ position and the SDSS position ($\lesssim 0.3\arcsec$) shows that the physical separation between the \hbox{X-ray} point-source and the optical centroid is less than $50$ pc at $z=0.002$, which further reduces the likelihood of J115851+4243.2 being an off-nuclear ULX. 
The luminous nuclear mid-IR emission based on the {\it WISE} photometry, and the old stellar population suggested by the optical spectrum of J115851+4243.2,  also support the presence of a heavily obscured AGN (see \S\ref{subsec:obscuration}). Therefore, we argue that J115851+4243.2 is indeed powered by accretion onto the central massive BH.

\paragraph{J120711+3348.5 and J121358+2936.1} %B2 & WAS49
J120711+3348.5 and J121358+2936.1 are both broad-line AGNs with many similar properties, including redshifts, $W1-W2$ colors, and $M_\bullet$.
While the \nus\ photon counts for both objects are limited, J121358+2936.1 has a higher flux in the \nus\ hard band than J120711+3348.5. 
For J120711+3348.5, we examine the existing \swiftxrt\ data and find it has only 10 photon counts above $>2$ keV. Thus, we cannot reliably constrain its obscuring column density. 
For J121358+2936.1, we fit the publicly available \chandra\ data with an absorbed power-law and found the best-fit model has a moderate obscuring column density ($N_{\rm H}\lesssim 10^{22}$ cm$^{-2}$). 

\paragraph{J134934--3025.5} For J134934--3025.5, the \xmm\ archival data has only 20 photon counts in the 2--10 keV band of EPIC-pn, which is not sufficient for spectral fitting. We estimate its $L_{2-10\mathrm{keV}}$ to be $\approx 2.9 \times 10^{42}$ erg s$^{-1}$ based on its {\it XMM}-Newton photon count rate, assuming an AGN photon index of 1.8 due to the limited photon counts. 
In this luminosity range, J134934--3025.5 is considered to be a {\it bona-fide} \hbox{X-ray} AGN. However, its optical spectrum exhibits a significant 
$4000$~\AA\ break and absorption lines in  Ca H, K and $\mathrm{H}\alpha$. This strongly suggests that J134934--3025.5 is a quiescent galaxy. As we have discussed at length in \S\ref{subsec:xbong}, future \hbox{X-ray} follow-up observations are required to determine the reason for the lack of optical emission lines in J134934--3025.5. 

\paragraph{J223654+3423.5}
J223654+3423.5 is a faint emission-line galaxy. 
Notably, its soft \hbox{X-ray} luminosity ($L_{2-10\mathrm{keV}} = 3.8\times10^{41}$ \ergs\ , estimated based on the \chandra\ observation with a fixed photon index of 1.8 using {\tt PIMMS}) is slightly lower than the $10^{42}$ \ergs\ limit for empirically separating AGNs from ULXs. While it is still possible for J223654+3423.5 to be an extremely luminous ULX, we note that ULXs with luminosities similar to that of J223654+3423.5 are likely to be powered by accretion onto off-nuclear mBHs \citep[e.g.,][]{walt14,mukh15}. Without high-resolution spatial information on the source of \hbox{X-ray} emission, the distinction between the mBH-powered ULX and low-mass AGN becomes ambiguous. However, this should not affect the primary objective of this work of searching for accreting mBHs using hard \hbox{X-ray} observations. Moreover, the \nus\ soft-band luminosity is $\approx 6$ times higher than that for the \chandra\ observation for J223654+3423.5, suggesting that the $L_{2-10\mathrm{keV}}$ is more luminous than $10^{42}$ \ergs\ during the time of the \nus\ observation. Therefore, we consider J223654+3423.5 to be a low-mass AGN powered by an mBH similar to the other objects in our sample.
We note that the separation between the \nus\ and \chandra\ observation dates are relatively short (56 days), but the origin of the variability of J223654+3423.5 could not be determined with the currently available data.

\section{Notes on additional soft X-ray observations not used in this work}
In \S\ref{subsec:softxray} and Table 4, we report that there are six sources in our sample with more than one soft X-ray observations. After careful considerations these observations were not used in this work. Here we present the details of these observations.
\paragraph{J023229+2023.7}
The \xmm\ PN data of obsdIDs 0604210201 and 0604210201 both suffer from significant high energy (10--12 keV) background flaring for more than 50\% of the observed duration, and both observations have less than 10~ks effective PN exposure time. Additionally, the source is located near the PN chip gap for both obsIDs. While the less-sensitive MOS detectors were not as severely affected by the flaring background, J023229+2023.7 is located on the MOS1 CCD that has been permanently shut off. The data on MOS2 was taken under ``timing mode'' for both obsIDs and we could not extract useful spectral information for J023229+2023.7. Therefore, we only adopt the \nus\ data for constraining the AGN X-ray property.

% and thus the only available data is from MOS2. We compare the best-fit \nus\ model with the MOS2 data and find that the MOS2 spectrum has substantial excess at $<5$~keV, suggesting that the \xmm\ spectrum might have significant contamination from the extended soft X-ray emission. We also jointly fitted the \xmm\ and \nus\ spectra with the original model but we cannot find statistically acceptable results. This suggests that the soft X-ray component requires a more complex model. However, the limited \xmm\ photon counts ($\sim 80$) prevents us from exploring further and thus we only adopt the \nus\ data for constraining the AGN X-ray property. 

\paragraph{J032459--0256.2}
We extract the EPIC PN spectrum from for the \xmm\ observation obsID 0405240201 using the similar approach described in \S\ref{subsec:softxray}. We jointly fit the \xmm\ data and the \nus\ data following the description of this source given in Appendix A and find that the best-fit parameters to be within the uncertainty range of the joint \swiftxrt\ and \nus\ fit. The best-fit X-ray luminosity is $L_{2-10\mathrm{keV}} = 2.3\times10^{41}$ erg s$^{-1}$ , which is slightly less than the result based on the \swiftxrt\ and \nus\ data ($L_{2-10\mathrm{keV}} = 3.3\times10^{41}$ erg s$^{-1}$). For this work, we choose to use the result based on the \swiftxrt\ and \nus\ data because its 3--8 keV X-ray flux is closer to that of the \nus\ data. 

\paragraph{J065318+7424.8}
The \xmm\ observation obsID 0061540101 has only $\sim 7$~ks background-filtered exposure time thus no useful spectral information is available. The \xmm\ data 0144230101 has $\sim 30$ ks exposure time, but the \xmm\ image appears to be  extended. We compare the spectrum of 0144230101 with the best-fit model based on the $\sim 70$ ks \chandra\ data and find that 0144230101 has significant excess below $3$ keV, suggesting the \xmm\ spectrum might be contaminated by the extended soft X-ray emission. To avoid large uncertainties due to the requirement of an additional soft X-ray component and the smaller photon counts of the \xmm\ observation, we use only the \chandra\ data for this work and do not consider the additional data from 0144230101. 

\paragraph{J101609--3329.6}
The exposure time of the additional \chandra\ data for J101609--3329.6 is only $\sim 10$~ks and the photon counts are too low to provide useful spectral constraints compared to the $\sim 80$~ks \xmm\ observation adopted for the main analysis. 

\paragraph{J115851+4243.2}
As described in Appendix A, there are multiple \chandra\ point sources within the $\sim 10\arcsec$ \xmm\ spectrum extraction region for the \xmm\ data 074404301. To avoid contaminations from these off-nuclear sources we do not make use of the \xmm\ observation. 

\paragraph{J223654+3423.5}
This source has three different 10ks \chandra\ observations and none of them have sufficient photon counts for spectral analysis. For the main article, we choose obsID 17570 because the flux of this obsID is the closest to that of the \nus\ observation. Further investigating the nature of the X-ray variability of J223654+3423.5 would require additional X-ray observations.

\section{Notes on the \textit{S\lowercase{wift}}/BAT low-mass galaxies}\label{appendix:bat_k11}
Here we briefly summarize the properties of the \swiftbat\ low-mass AGNs discussed in this paper. K11 selected local ($z\lesssim 0.05$) AGNs from the 22-month and 58-month \swiftbat\ catalogs in the northern sky (DEC $> -25$ deg). For galaxies without SDSS photometry, K11 observed them using the Kitt Peak 4-meter telescope with the same filters as those of SDSS. The nuclear contribution to the photometry was then removed for each galaxy using surface-brightness profile fitting methods, and the host-galaxy Petrosian magnitudes were measured using an automated pipeline identical to the SDSS one. The stellar mass for each galaxy was measured using the {\sc kcorrect} package.
Of the 185 galaxies in K11, 38 of them are ``low-mass'' galaxies with $M_\star<10^{10} M_\sun$. For this work, we discard the six galaxies with more than $50\%$ AGN contribution at {\it r-}band to avoid the selection of galaxies with uncertain stellar masses. The median {\it r-}band absolute magnitude for the rest of the 32 low-mass AGNs is $-20.11$, which is only slightly lower than the median of our \nus\ sample ($-20.03$).
To test whether the $M_\star$ estimated in K11 is systematically different to the $M_\star$ of our \nus\ sample estimated using the SED-fitting approach, we obtain the optical to mid-IR photometry for K11 objects within the SDSS footprint and use the SED-fitting method described in \S\ref{subsec:sedfitting} to recalculate their $M_\star$.
We find that the $M_\star$ measured using our SED-fitting approach is slightly lower than the K11 $M_\star$ by a median value of $0.09$ dex. With the much larger $\approx 0.3$ dex uncertainty caused by the stellar population synthesis model degeneracy 
\citep[e.g.][]{conr09mstar}, we consider the $M_\star$ of the K11 low-mass AGNs and our \nus\ sample to be directly comparable.

For the hard \hbox{X-ray} luminosity, we match the K11 AGNs with the \swiftbat\ 70-month catalog and convert the \hbox{$14-195$} keV luminosity provided in the 70-month catalog to \hbox{$10-40$} keV luminosity assuming a typical AGN spectrum with a photon-index of $\Gamma=1.8$. The correction is $\approx 0.4$ dex for the redshift range of the K11 sample. The column densities for the K11 AGNs are culled from C. Ricci et al. (in preparation), which analyzes the soft X-ray spectra of all BAT AGNs using archival data \citep[also see][]{ricc15}. The key properties of the K11 low-mass AGNs are summarized in Table~\ref{tab:bat}.

\newpage
\onecolumngrid
\begin{deluxetable}{lccchhcccch}
\tablecaption{Key properties of \swiftbat\ low mass galaxies selected from Koss et al. (2011)\label{tab:bat}}
\tablehead{
\colhead{Name} &
\colhead{RA} &
\colhead{DEC} &
\colhead{{\it z}} &
\nocolhead{u-r} &
\nocolhead{g-r} &
\colhead{$\log M_\star$} &
\colhead{\swiftbat\ ID} &
\colhead{$\log L_{10-40\mathrm{keV}}$} &
\colhead{$\log N_{\rm H}$} &
\colhead{} \\
\colhead{} & \colhead{(J2000)} & \colhead{(J2000)} & \colhead{} & \nocolhead{} &\nocolhead{(mag)} &
\colhead{($\log M_\sun$)} & \colhead{} & \colhead{($\log$ erg s$^{-1}$)} & \colhead{($\log$ cm$^{-2}$)} & \colhead{}
}
\colnumbers
\startdata
Mrk 352                 & 14.972 & 31.8269 & 0.0149 & \nodata & 0.67 & 9.65 & SWIFT J0059.4+3150 & 42.87 & 20.0 & W09 \\
2MASX J03534246+3714077 & 58.427 & 37.235 & 0.0183 & 2.34 & 0.49 & 9.9 & SWIFT J0353.7+3711 & 42.84 & 22.34 & \nodata \\
2MASX J05054575-2351139 & 76.4405 & -23.8539 & 0.035 & 1.87 & 0.67 & 9.13 & SWIFT J0505.8-2351 & 43.84 & 22.85 & W09 \\
MCG -05-14-012          & 85.8873 & -27.6514 & 0.0099 & 1.73 & 0.63 & 9.66 & SWIFT J0543.9-2749 & 42.4 & 20.0 & \nodata \\
2MASX J06411806+3249313 & 100.3252 & 32.8254 & 0.047 & 1.95 & 0.61 & 9.94 & SWIFT J0641.3+3257 & 43.83 & 23.3 & W09 \\
Mrk 1210                & 121.0244 & 5.1138 & 0.0135 & 1.62 & 0.65 & 9.89 & SWIFT J0804.2+0507 & 43.09 & 23.0 & B99 \\
Mrk 18                  & 135.493 & 60.152 & 0.0111 & 1.75 & 0.63 & 9.57 & SWIFT J0902.0+6007 & 42.06 & 23.3 & W09 \\
2MASX J09043699+5536025 & 136.154 & 55.6008 & 0.037 & 1.8 & 0.44 & 9.76 & SWIFT J0904.3+5538 & 43.37 & 20.78 & W09 \\
2MASX J09112999+4528060 & 137.8749 & 45.4683 & 0.0268 & 2.38 & 0.79 & 9.76 & SWIFT J0911.2+4533 & 43.21 & 23.48 & W09 \\
IC 2461                 & 139.992 & 37.191 & 0.0075 & 3.43 & 0.67 & 9.54 & SWIFT J0920.1+3712 & 42.13 & 22.85 & N09 \\
Mrk 110                 & 141.3036 & 52.2863 & 0.0353 & 1.99 & 0.63 & 9.9 & SWIFT J0925.0+5218 & 43.95 & 20.3 & W09 \\
CGCG 122-055            & 145.52 & 23.6853 & 0.0214 & 2.67 & 0.69 & 9.94 & SWIFT J0942.2+2344 & 42.65 & 20.1 & \nodata \\
NGC 3079 				& 150.4908 & 55.6798 & 0.0037 & 1.57 & 0.63 & 9.98 & SWIFT J1001.7+5543 & 41.6 & 22.3 & B99 \\
NGC 3227 				& 155.8774 & 19.8651 & 0.0039 & 2.73 & 0.84 & 9.98 & SWIFT J1023.5+1952 & 42.29 & 22.3 & W09 \\
ARP 151 				& 171.4007 & 54.3825 & 0.0211 & 2.46 & 0.93 & 9.71 & SWIFT J1125.6+5423 & 43.02 & 21.7 & V13 \\
NGC 3718 				& 173.1452 & 53.0679 & 0.0033 & 2.46 & 0.7 & 9.98 & SWIFT J1132.7+5301 & 41.2 & 22.0 & \nodata \\
MCG+10-17-061 			& 176.3881 & 58.9781 & 0.0099 & 2.12 & 0.57 & 9.8 & SWIFT 1145.2+5905 & 42.55 & 22.9 & V13 \\
NGC 4051 				& 180.7901 & 44.5313 & 0.0023 & 1.78 & 0.52 & 9.44 & SWIFT J1203.0+4433 & 41.5 & 20.0 & Tartarus \\
NGC 4102 				& 181.5963 & 52.7109 & 0.0028 & 2.4 & 0.72 & 9.68 & SWIFT J1206.2+5243 & 41.26 & 24.48 & V13 \\
NGC 4138 				& 182.3741 & 43.6853 & 0.003 & 2.38 & 0.71 & 9.61 & SWIFT J1209.4+4340 & 41.37 & 23.0 & V13 \\
Mrk 50 					& 185.8506 & 2.6791 & 0.0234 & 2.38 & 0.72 & 9.9 & SWIFT J1223.7+0238 & 43.17 & 20.0 & V13 \\
NGC 4395 				& 186.4538 & 33.5468 & 0.0011 & 1.18 & 0.3 & 8.28 & SWIFT J1202.5+3332 & 40.5 & 21.3 & Tratarus \\
ESO 506-G027 			& 189.7275 & -27.3078 & 0.025 & 3.88 & 0.66 & 9.84 & SWIFT J1238.9-2720 & 43.72 & 23.9 & W09 \\
SBS 1301+540 			& 195.9978 & 53.7917 & 0.0299 & 2.2 & 0.57 & 9.79 & SWIFT J1303.8+5345 & 43.52 & 22.3 & V13 \\
NGC 5273 				& 205.5347 & 35.6542 & 0.0035 & 3.34 & 0.72 & 9.64 & SWIFT J1341.9+3537 & 41.06 & 20.0 & V13 \\
UM 614 					& 207.4701 & 2.0791 & 0.0327 & 0.5 & 0.68 & 9.99 & SWIFT J1349.7+0209 & 43.3 & 21.0 & V13 \\
Mrk 464 				& 208.973 & 38.5746 & 0.0501 & \nodata & 0.48 & 9.67 & SWIFT J1356.1+3832 & 43.8 & 24.0 & V13 \\
Mrk 477 				& 220.1587 & 53.5044 & 0.0377 & 1.24 & 0.05 & 9.87 & SWIFT J1441.4+5341 & 43.36 & 22.95 & B99 \\
NGC 5995 				& 237.104 & -13.7578 & 0.0252 & 1.95 & 0.75 & 9.89 & SWIFT J1548.5-1344 & 43.52 & 22.0 & Tartarus \\
CGCG 300-062            & 265.8225 & 62.8392 & 0.033 & 2.01 & 0.61 & 9.9 & SWIFT J1743.4+6253 & 43.08 & 23.0 & \nodata \\
2MASX J21355399+4728217 & 323.975 & 47.4727 & 0.025 & 2.2 & 0.72 & 9.41 & SWIFT J2156.1+4728 & 43.27 & 21.6 & W09 \\
KAZ 320 				& 344.8871 & 24.9182 & 0.0345 & 0.94 & 0.2 & 9.7 & SWIFT J2259.7+2458 & 43.57 & 20.0 & \nodata \\
\enddata
\tablecomments{
Column 1 : Source name. Columns 2-3 : \swiftbat\ RA/DEC (J2000). Column 4 : redshift. Column 5 : stellar mass. Column 6 : \swiftbat\ 70-month ID \citep{cat_swiftbat70}. Column 7 : $L_{10-40\mathrm{keV}}$ calculated from $L_{14-195 \mathrm{keV}}$ of the \swiftbat\ 70-month catalog (see Appendix B). Column 8 : Intrinsic $N_{\rm H}$, see Appendix B for details.}
\end{deluxetable}
\normalsize

%\bibliography{../../../TEX/BIB/nustar_dwarf_16}

\end{document}